\documentclass[prl,twocolumn,superscriptaddress,longbibliography,amsmath,amssymb,aps,floatfix]{revtex4-1}

\usepackage{graphicx}% Include figure files

\usepackage{siunitx}
\usepackage{mathrsfs}

\DeclareSIUnit\px{px}
\newcommand{\asp}{\lambda}
\newcommand{\decay}{\mu}
\newcommand{\ve}[1]{\ensuremath{\mbox{\boldmath$#1$}}}
\newcommand{\ma}[1]{\ensuremath{\mathbb{#1}}}

\usepackage{ragged2e}

\usepackage{xcolor}

\begin{document}

\title{Inertia induces strong orientation fluctuations of non-spherical atmospheric particles}

\author{T. Bhowmick}
\thanks{These authors contributed equally to this work}
\affiliation{Max Planck Institute for Dynamics and Self-Organization, G\"ottingen, D-37077 Germany}
\affiliation{Institute for the Dynamics of Complex Systems, University of G\"ottingen, Friedrich-Hund-Platz 1, G\"ottingen, D-37077 Germany}
\author{J. Seesing}
\thanks{These authors contributed equally to this work}
\affiliation{Max Planck Institute for Dynamics and Self-Organization, G\"ottingen, D-37077 Germany}
\author{K. Gustavsson}
\thanks{These authors contributed equally to this work}
\affiliation{Department of Physics, Gothenburg University, Gothenburg, SE-40530 Sweden}
\author{J. Guettler}
\affiliation{Max Planck Institute for Dynamics and Self-Organization, G\"ottingen, D-37077 Germany}
\author{Y. Wang}
\affiliation{Max Planck Institute for Dynamics and Self-Organization, G\"ottingen, D-37077 Germany}
\author{A. Pumir}
\affiliation{Laboratoire de Physique, ENS de Lyon, Universit\'e de Lyon 1 and CNRS, Lyon, F-69007 France}
\affiliation{Max Planck Institute for Dynamics and Self-Organization, G\"ottingen, D-37077 Germany}
\author{B. Mehlig}
\affiliation{Department of Physics, Gothenburg University, Gothenburg, SE-40530 Sweden}
\author{G. Bagheri}
\email[Corresponding author:\\]{gholamhossein.bagheri@ds.mpg.de}
\affiliation{Max Planck Institute for Dynamics and Self-Organization, G\"ottingen, D-37077 Germany}

\date{\today}

\begin{abstract}
The orientation of non-spherical particles in the atmosphere, such as volcanic ash and ice crystals, influences their residence times, and the radiative properties of the atmosphere. Here, we demonstrate experimentally that the orientation of heavy submillimeter spheroids settling in still air exhibits decaying oscillations, whereas it relaxes monotonically in liquids. Theoretical analysis shows that these oscillations are due to particle inertia, caused by the large particle-fluid mass-density ratio. This effect must be accounted for to model solid particles in the atmosphere.
\end{abstract}


\maketitle

\emph{Introduction.}-- The transport, dispersion, and settling of
volcanic ash~\cite{Bagheri_Bonadonna_2016, Rossi_2021}, microplastic particles~\cite{Allen:2019,Zhang:2020},
and ice crystals in cold atmospheric clouds~\cite{Cox_Arnold_1979,Pruppacher_Klett_2010,Garrett_2015,Wang_2013}
have significant environmental impact.
These non-spherical particles are subject to
gravity, as well as viscous and inertial hydrodynamic forces and torques.
An essential parameter characterising the latter is the particle Reynolds number, defined by ${\rm Re}_{\rm p} = a v_g /\nu$,
where $a$ is the linear size of the particle, $v_g$ is its settling speed, and $\nu$ is the kinematic viscosity of air.
In general, the transport of non-spherical particles depends strongly
on their
angular dynamics~\cite{Cox_1965,Happel_Brenner_1983,Khayat_Cox_1989}, which directly
affects the settling speed~\cite{Pruppacher_Klett_2010,Voth_Soldati_2017,Bagheri_Bonadonna_2016,Roy_2019}. This,
in turn, determines its residence times and dispersion ranges in the atmosphere.
The settling speed influences, for instance, how far microplastic, dust and volcanic ash can be transported away from a  source, or how much time an ice crystal spends in a cloud~\cite{Allen:2019,Bagheri_Bonadonna_2016,Baran_2012,Rossi_2021}.
In addition, the angular dynamics determines the volume
swept out. Together with the settling speed, this volume is a  key parameter determining particle-particle collision rates \cite{pumir2016collisional}, e.g. relevant for the formation of aggregates of ice particles in clouds  \cite{Pinsky:1998,siewert2014collision,Sheikh_2022,Gavze_2022} or volcanic ash  \cite{Rossi_2021}.
The particle orientation also has a direct impact on the absorption and scattering of radiation by the atmosphere,  which affects albedo of atmospheric clouds~
\cite{Krotkov_1999,Mishchenko_1996,Noel_Chepfer_2004}, an effect still not understood quantitatively, despite its importance~\cite{Noel_Sassen_2005,Westbrook_2010,Noel_Chepfer_2010,Marshak:2018,Gustavsson_2021}.

Many studies investigated the drag and stable orientation of non-spherical particles settling in viscous liquids  at rest~\cite{willmarth1964steady,Jayaweera_Cottis_1969,swaminathan2006sedimentation,Roy_2019,Esteban_2020,Cabrera_2022}, where the ratio $\mathscr{R} = \rho_{\rm p}/\rho_{\rm f}$ between the particle-mass density $\rho_{\rm p}$ and the fluid-mass density $\rho_{\rm f}$  is close to unity.  
When the particle Reynolds number ${\rm Re}_{\rm p}$ is larger than unity, the particle orientation aligns rapidly and monotonically, causing the particle to settle with its broad side down~\cite{Khayat_Cox_1989,Gustavsson_2019}.
The angular dynamics becomes unsteady at larger ${\rm Re}_{\rm p}$. Settling particles in liquids, for example,  exhibit a rich variety of motion patterns at ${\rm Re}_{\rm p}\sim 100$~\cite{auguste2013falling}.
In air with  $\mathscr{R}\sim 1000$, the angular dynamics of thin settling disks
exhibits bistability~\cite{coletti2023} at ${\rm Re}_{\rm p}\sim 50$. 
One expects that particle inertia plays an important role in explaining this qualitative difference, since $\mathscr{R}$ is very large.

We remark that at  much higher ${\rm Re}_{\rm p}$, the angular dynamics of non-spherical particles
settling in still air becomes unstable~\cite{Bagheri_Bonadonna_2016}.
Vortex shedding causes the characteristic fluttering first considered by Maxwell, see Ref.~\cite{pesavento2004falling} and references cited there.

In turbulence, where
fluid-velocity gradients give rise to additional torques, most experiments concern particles with $\mathscr{R}\approx 1$~\cite{Kramel,Voth_Soldati_2017,Lopez_Guazzelli_2017,Bounoua_2018,Esteban_2020,Roy_2019,Roy23}.
Only a few experiments with turbulent fibre suspensions were carried out in  air~\cite{newsom1998orientational,QI_2012,Kuperman:2019}.
For the tumbling of heavy fibres
in turbulence, particle inertia plays a
role~\cite{Einarsson_2014,Bounoua_2018},
but its effect remains to be quantified.

\begin{figure}[t]
\centering
\includegraphics[width=\columnwidth]{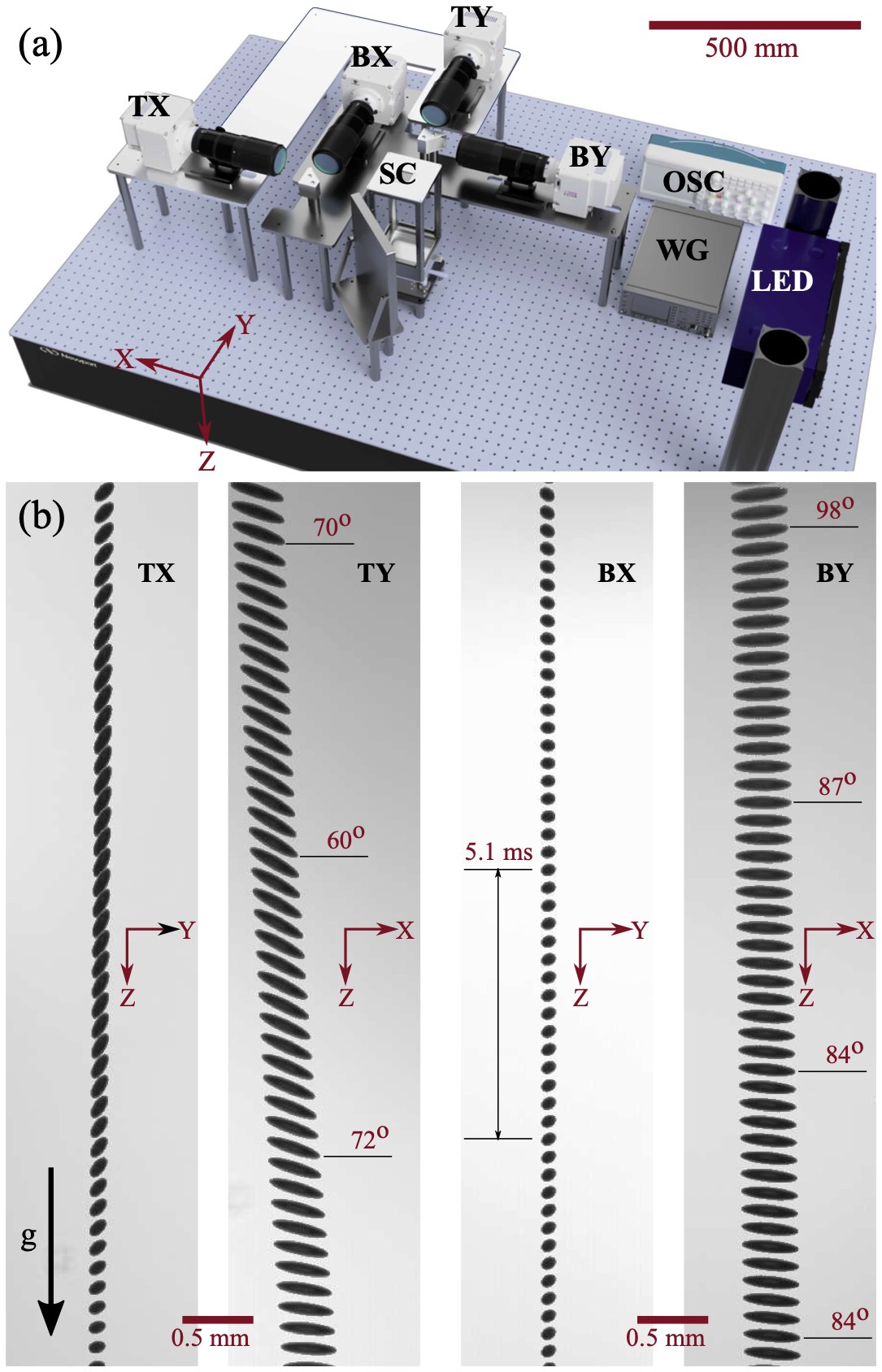}
\caption{\label{fig:Fig1}
Experimental setup.
(a) Optical table with top cameras (TX and TY), and bottom cameras (BX and BY) named after the shown coordinate system ($Z$ is the direction of gravity $\ve g$), the settling chamber (SC), the synchronized pulsed LED unit (LED), controlled with a waveform generator (WG), and the oscilloscope (OSC).
(b) Snapshots of a settling prolate spheroid recorded at 2932 frames per second.
The tilt-angle -- the angle between the particle symmetry axis and gravity --  is shown  in $5.1 {\rm ms}$ intervals in units of degrees.
See Supplemental Material (SM)~\cite{sm} for  details.
}
\end{figure}

Experimentally, it remains a challenge  to study the inertial
angular dynamics of particles that
settle very rapidly.
One needs  precise particle tracking over long periods of time, high-magnification imaging,  a particle-release mechanism that does not set the fluid in motion,
and the container must be large enough to avoid spurious interaction with the wall
\cite{swaminathan2006sedimentation}.
We developed  a new experimental apparatus that overcomes these challenges
 [Fig.~\ref{fig:Fig1}(a)]. Here we report on the first measurements, showing
 that the particle
orientation exhibits
characteristic oscillations with time scales comparable to those
of atmospheric turbulence. The particles --
with Reynolds numbers between $2$ and $35$ -- do eventually align
so that they settle with their broad side down.
We explain  that the oscillations are induced by particle inertia, and conclude that particle inertia can significantly enhance the extent to which turbulence randomises particle orientation.

\emph{Experiments.}-- The setup consists of an air-filled settling chamber (SC) with a novel particle injector, and two high-speed camera pairs (TX, TY) and (BX, BY), synchronised with a high-intensity pulsed LED array [Fig.~\ref{fig:Fig1}(a)].
Each camera  images a fall distance of $\sim\SI{30}{\milli\meter}$ at a nominal resolution of \SI{6.75}{\micro\meter\per\px},
The apparatus allows us to  image individual solid $0.1-\SI{5}{\milli\meter}$-particles
settling in quiescent air.
See Supplemental Material (SM)~\cite{sm} for a complete description of the setup.

\begin{table}
\caption{\label{tab:sd}%
Size groups of particles studied. Parameters: aspect ratio $\asp = a_\|/a_\perp$ ($2a_\parallel$ is the particle length along its symmetry axis,
and $2a_\perp$ is its diameter); volume  $V_{\rm p}$; Reynolds number
${\rm Re}_{\rm p}$ (using $a={\rm max}\{a_\parallel,a_\perp\}$ and  the observed
settling speed);  Stokes time  $\tau_{\rm p} = ({2\rho_{\rm p}}/{9\rho_{\rm f}}) a_\perp a_\|/\nu$, where $\rho_{\rm p}$ and $\rho_{\rm f}$ are the mass densities of the particle and the
fluid, and $\nu=1.5\times 10^{-5}$m$^2$/s is the kinematic viscosity of air.}
\begin{ruledtabular}
\begin{tabular}{ccccccc}
        Group & $\asp$ & $2a_\parallel$ [\SI{}{\micro\meter}] & $2a_\perp$ [\SI{}{\micro\meter}] & $V_{\rm p}$ [\SI{}{{\milli}\meter\cubed}] & ${\rm Re}_{\rm p}$ & $\tau_{\rm p}$ [\SI{}{\milli\second}] \\
    \colrule
        I & 0.2{0} & 47.9 & 239.4 & {\SI{1.44e-3}{}} & 2.8 & 42{.0} \\
        I & 0.5{0} & 88.2 & 176.4 & {\SI{1.44e-3}{}} & 2.5 & 57{.0} \\
        I & 0.8{0} & 120.6 & 150.8 & {\SI{1.44e-3}{}} & 2.4 & 66.7 \\
        I & 1{.00} & 140{.0} & 140{.0} & {\SI{1.44e-3}{}} & 2.2 & 71.8 \\
        I & 1.25 & 162{.0} & 130{.0} & {\SI{1.44e-3}{}} & 2.6 & 77.2 \\
        I & 2{.00} & 222.2 & 111{.0} & {\SI{1.44e-3}{}} & 3.3 & 90.4 \\
        I & 5{.00} & 410{.0} & 81.8 & {\SI{1.44e-3}{}} & 5{.0} & 122.9 \\
    \colrule
        II & 0.25 & 65.5 & 261.9 & {\SI{2.35e-3}{}} & 3.8 & 62.9 \\
        II & 4{.00} & 399.4 & 99.9 & {\SI{2.08e-3}{}} & 5.9 & 146.3 \\
    \colrule
        III & 0.25 & 150{.0} & 600{.0} & {\SI{28.28e-3}{}} & 22.5 & 329.9 \\
        III & 4{.00} & 876.9 & 219.2 & {\SI{22.07e-3}{}} & 34.3 & 704.6 \\
\end{tabular}
\end{ruledtabular}
\end{table}

We used the Photonic Professional GT 3D printer (Nanoscribe
GmbH) with sub-micrometer printing resolution to print submillimetre-sized spheroids with mass density $\rho_{\rm p} = \SI{1200}{\kg\per\meter\cubed}$~\cite{Liu_2018}.
 Three size groups of spheroids were produced (Table~\ref{tab:sd}), with different aspect ratios $\lambda=a_\parallel/a_\perp$
  ($2a_\parallel$ is the length of the  particle symmetry-axis, and  $2a_\perp$ is the perpendicular diameter).
Three-dimensional scans of the particle shape using a 3D laser-scanning microscope (VK-X200K, Keyence) show that the unevenness in the surface features of the spheroids is negligible (Fig.~S2 in the SM~\cite{sm}).
In total, we carried out between $9$ and $22$ measurements per particle shape and size, resulting in a total of $170$
successful experimental runs
where the particle was in sharp focus for all four cameras.
Fig.~\ref{fig:Fig1}(b) shows recorded images of a prolate spheroid  ($2a_\parallel = \SI{410}{\micro\meter}$, $2a_\perp = \SI{82}{\micro\meter}$) as it falls in the settling chamber.
Figs.~S3-S4~\cite{sm} contain recorded images of all particles in Table~\ref{tab:sd}.

Fig.~\ref{fig:Fig1}(b) indicates
that the orientation of the particle oscillates as it settles, in sharp contrast with
previous experiments
in liquids,
where the alignment is monotonic~\cite{Kramel,Roy23,Lopez_Guazzelli_2017,Gustavsson_2019,Roy_2019}. To explain the oscillations observed here, we developed a theoretical model that includes the effect of particle inertia.

\begin{figure}[t]
\includegraphics[width=\columnwidth]{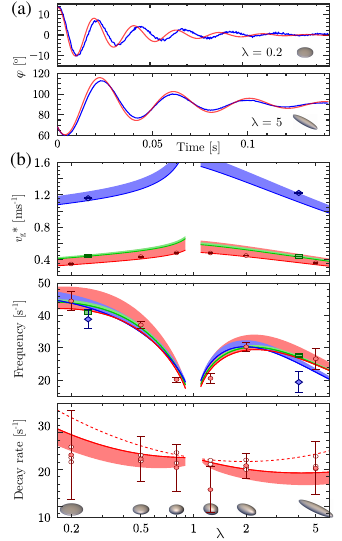}
\caption{\label{fig:Fig2}
Comparison between experiments and theory.
(a) Time evolution of tilt-angle $\varphi$ from experiment (blue) and its model prediction (red) for spheroids with aspect ratios $\asp=0.2$ and $5$ from group I (Table~\ref{tab:sd}).
(b) Steady-state settling speed $v^\ast_g$, frequency, and decay rate of the tilt-angle oscillations as functions of the aspect ratio $\asp$.
Markers show averages obtained for all experiments with error bars indicating 95\% confidence bounds for groups
I~(red/$\circ$), II~(green/$\Box$), and III~(blue/$\diamond$) (Table~\ref{tab:sd}).
Solid lines show large-time asymptotes from a linear-stability analysis of
the model described in Appendix~A.
Shaded regions indicate how much the theoretical predictions change as the measured settling speed varies along the particle trajectory.
See Appendices A and B for details.
Dashed lines show results of a linear-stability analysis of a harmonic-oscillator approximation, Eq.~(\ref{eq:linear}) in Appendix~D}.
\end{figure}

\emph{Model.}-- Since the particle-to-fluid mass-density ratio $\mathscr{R}$  is
large, one suspects that the observed oscillations in the angular
dynamics are due to particle inertia. This expectation is consistent
with a recent theoretical study~\cite{Gustavsson_2021} of tilt-angle fluctuations
of small particles in turbulence.
However, we cannot directly apply the model from Ref.~\cite{Gustavsson_2021}, because that model was derived
assuming ${Re}_{\rm p} \ll 1$, whereas our particles have ${\rm Re}_{\rm p}$ of
order~$10$.
Therefore, we developed a theoretical description
that extends the validity of the ${\rm Re}_{\rm p}\ll 1$-model.
Specifically, we modified the fluid-inertia contributions
to the hydrodynamic force and torque by introducing
two scalar functions $C_F$ and $C_T$ that depend on the settling speed, determined using ab-initio simulations of spheroids fixed in a uniform flow~\cite{ouchene2016new,andersson2019forces,ouchene2020numerical,frohlich2020correlations,Jiang_2021,Roy23}. Details are given in Appendix~A.
As shown below, the new model accurately describes the experimental results, and highlights the key significance of particle inertia.
The  model has three non-dimensional parameters (Appendix~C): the aspect ratio $\asp = a_\|/a_\perp$, the non-dimensional particle volume, $\mathscr{V} = g V_{\rm p}/\nu^2$, where
$g$ is the gravitational acceleration,
$V_{\rm p} = (4 \pi/3) a_\perp^2 a_\|$ is the particle volume, and the mass-density ratio $\mathscr{R}=\rho_{\rm p}/\rho_{\rm f}$.
These parameters
determine the steady-state settling speed $v_g^\ast$, and thus ${\rm Re}_{\rm p}$. For small settling speeds,
${\rm Re}_{\rm p} \approx (1/6 \pi) \mathscr{R} \mathscr{V}$, up to a $\asp$-dependent
factor of order unity.
For oblate disks, particle inertia can be parameterised by a non-dimensional inertia ratio,~$J^\ast$.
It is related to our non-dimensional parameters by $J^* = (\pi/64) \mathscr{R} \asp$ (Appendix~C).
Our ${J}^*$-values are at least two orders of magnitude larger than for disks settling in water~\cite{willmarth1964steady,Ern_2012,auguste2013falling}, and larger than in~\cite{coletti2023} by a factor $\sim 2$.

\emph{Results.}-- The experimental results are compared with model predictions in Fig.~\ref{fig:Fig2}.
Panel (a) shows
how the angle between the particle-symmetry vector and gravity (the tilt-angle $\varphi$)
decays to the
steady value $\varphi^\ast=0$ for disks and $\varphi^\ast = \tfrac{\pi}{2}$ for rods.
The decay is  oscillatory as opposed to the behavior in water~\cite{Roy_2019,Cabrera_2022}
where the dynamics is overdamped (particle inertia is negligible) and the decay is monotonic. This is the case when
 the  damping time $\tau_\omega$ of the angular velocity  is smaller than the decay time $\tau_\varphi \sim \nu/[v_g^\ast]^2$ of the tilt angle \cite{Gustavsson_2019}, which happens
when 
 $\mathscr{R}^3 \mathscr{V}^2 \ll1 $ (Appendix~C).
In the opposite limit, 
$\tau_\omega \gg \tau_\varphi$. So particle inertia lengthens the  transient.

Fig.~\ref{fig:Fig2}(b) demonstrates that the model captures the observed settling dynamics very well. The largest disagreement is in the decay rate, which is hard to measure
especially for nearly spherical particles.
The white markers  in the bottom panel  show selected experiments where the decay was best fitted by an exponential.
The smaller scatter of these data points, closer to the theory, suggests that systematic errors in extracting the data provide the most likely explanation of  differences between theory and experiments.
Fig.~\ref{fig:Fig2} reveals good agreement over the whole range of ${\rm Re}_{\rm p}$ and $\asp$ covered by our experiments.
We mention that differences between model and experiment are expected to grow
at larger ${\rm Re}_{\rm p}$,
because the determination of the inertial torque becomes less reliable
beyond ${\rm Re}_{\rm p} \approx 30$~\cite{Jiang_2021}.

\begin{figure}[t]
\centering
\includegraphics[width=\columnwidth]{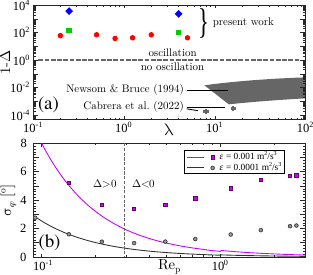}
\caption{\label{fig:Fig3}
(a) Bifurcation diagram.
The dashed horizontal line distinguishes decay without oscillations ($\Delta >0$) from decay with oscillations ($\Delta < 0$). 
Particles from group I in  Table~\ref{tab:sd} are shown as red/$\circ$, group II as green/$\Box$, group III as blue/$\diamond$. Particles from
 Ref.~\cite{Cabrera_2022} as $+$, and fibers from Ref.~\cite{Newsom_Bruce_1994} as a gray region.
We approximated cylindrical fibers as slender prolate spheroids and estimated $\Delta$ by setting $C_T = C_F = 1$, since the corresponding ${\rm Re}_{\rm p}$ are very small. (b) Standard deviation of tilt-angle fluctuations for spheroids  settling in weakly turbulent air with dissipation rate $\varepsilon$, as a function of ${\rm Re}_{\rm p}$.
Shown are simulation results of the model described in Appendix~E in symbols,
for spheroids with $\asp = 0.2$ but different volumes, $V_{\rm p}$.
Solid lines correspond to  the overdamped approximation, neglecting particle inertia (Eq.~(S6) in  SM~\cite{sm}).
The harmonic-oscillator bifurcation ($\Delta = 0$) is shown as a dashed  vertical line.
}
\end{figure}

To develop a qualitative understanding of the oscillations, we simplified the model further, assuming that $\delta\varphi = \varphi-\varphi^\ast$ remains small and that the settling speed is large.
In this limit we obtain a harmonic-oscillator equation for  $\delta\varphi$, namely $\delta \ddot \varphi +\delta \dot \varphi + ({V}_g^\ast)^2 C_T {|} h(\asp){|}  \mathscr{R}^3 \mathscr{V}^2 \delta\varphi =0$ in non-dimensional units (Appendix~C). In particular, $V_g^\ast$ is a non-dimensional settling speed $\propto v_g^\ast/(g\tau_{\rm p})$, and $h(\asp)$ is a shape-dependent function, shown in Fig.~S6.
For ${\rm Re}_{\rm p}\ll 1$ and $|\delta\varphi|\ll 1$, this equation simplifies to the form given in Refs.~\cite{Klett_1995,Gustavsson_2019,Gustavsson_2021}.
Linear stability analysis of
the harmonic-oscillator approximation shows that the particles approach alignment
with an exponential decay,
with decay rate
$\decay_{\pm} = -\tfrac{1}{2} \pm \tfrac{1}{2} \sqrt{\Delta}$,
with discriminant $\Delta = 1- 4 ({V}_g^\ast)^2 C_T {|} h(\asp){|}  \mathscr{R}^3 \mathscr{V}^2$. The non-dimensional steady-state settling speed ${V}_g^\ast$ is of order unity.
For all  particles in our experiments, the values of ${|} h(\asp){|} \mathscr{R}^3\mathscr{V}^2$ were large enough to ensure that
$\Delta < 0$, as shown in Fig.~\ref{fig:Fig3}(a).
A qualitative change  occurs when  $\Delta $ becomes positive: then
the particle orientation relaxes without oscillation. The expression for $\Delta$ shows that this bifurcation 
cannot occur in the overdamped limit $\mathscr{R}^3 \mathscr{V}^2\ll1$. We conclude: 
the bifurcation is due to particle inertia.

Now consider the very slender fibres measured in Ref.~\cite{Newsom_Bruce_1994}.
The asymptotic form of ${|} h(\asp){|} $ for large $\asp$ implies that
${|} h(\asp){|}  \mathscr{R}^3 \mathscr{V}^2 \propto a_\perp^6$, disregarding factors of $\log\asp$.
It follows that, for $\mathscr{R} \sim 1000$, only fibres
with $a_\perp$ larger than
$\sim 20 \mu {\rm m}$ can oscillate. This explains why the fibres with diameters $\sim 10 \mu {\rm m}$ used in Ref.~\cite{Newsom_Bruce_1994}, represented by the gray region in Fig.~\ref{fig:Fig3}(a),
did not  oscillate.
We conclude that the angular dynamics of slender fibres in the atmosphere can be very
different from that of particles of moderate aspect ratios.
In case of very slender disks, ${|} h(\asp){|}  \approx 7 \times 10^{-5} \asp$, so
the parameter combination $| h(\asp ){|}\mathscr{R}^3\mathscr{V}^2$ depends on particle geometry as $(a_\perp \sqrt{\asp})^6$.
Our estimates indicate that oscillations are observable for thin disks when $a_\perp \sqrt{\asp}$ is larger than $\sim 20 \mu {\rm m}$.
This condition is very well fulfilled for the oblate particles in Table~\ref{tab:sd}, as well as for a large class of ice crystals~\cite{Pruppacher_Klett_2010}.
A potential shortcoming of this analysis is that the forces and torques acting on thin disks have not been thoroughly tested for values significantly smaller than $\asp \sim 0.1$~\cite{ouchene2020numerical,Jiang_2021}.

In turbulence, tilt-angle fluctuations are determined by balancing the inertial relaxation described above with the effect of turbulent fluid-velocity gradients which upset  alignment.
Typical Kolmogorov times for weak atmospheric turbulence, $\tau_{\rm K} \sim (\nu/\varepsilon)^{1/2} \sim 0.4$s to $0.04$s for the dissipation rate per unit mass
$\varepsilon$  from $10^{-4}$ m$^2$/s$^3$ to $10^{-2}$ m$^2$/s$^3$ \cite{Pruppacher_Klett_2010}
are of the same order as the
time scales in Fig.~\ref{fig:Fig2}, indicating that particle inertia can significantly increase the  randomising effect of turbulence,
just as for small spherical particles, where particle inertia matters most for Stokes numbers of order unity and larger
\cite{pumir2016collisional,Gus15a,brandt2022particle,Bec_2023}.
To describe the effect of turbulence, we added a stochastic forcing \cite{Bec_2023} to our model (Appendix~E).
Figure~\ref{fig:Fig3}(b) summarises the results. It shows  the  standard deviation $\sigma_\varphi$
of the tilt-angle fluctuations for small turbulent dissipation rate, obtained by simulations of  the model from Appendix~E. 
 Also shown is the overdamped approximation obtained by neglecting translational and rotational accelerations (Eq.~(S6) in the SM~\cite{sm}).
 In Fig.~\ref{fig:Fig3}(b), ${\rm Re}_{\rm p}$ was varied by changing the particle volume, keeping all other parameters the same.
 For small ${\rm Re}_{\rm p}$, simulations and overdamped theory agree, so particle inertia has no effect. 
As ${\rm Re}_{\rm p}$ grows, $\sigma_\varphi$ decreases at first, because the fluid-inertia torque aligns the particle more strongly as the settling speed increases.
At the same time, a difference between simulations and overdamped theory develops: particle inertia enhances the tilt-angle fluctuations.

At still larger ${\rm Re}_{\rm p}$, $\sigma_\varphi$ starts to increase again, forming a characteristic minimum.
This can be understood in terms of the harmonic-oscillator approximation:
the minimum in Fig.~\ref{fig:Fig3}(b) occurs at a critical value of ${\rm Re}_{\rm p}$ (details in SM~\cite{sm}) where $\Delta$ becomes negative,
indicating that the increase is due to the bifurcation described above, causing transient oscillations that result in larger angular fluctuations.
The bifurcation occurs when the Green's function  of the harmonic-oscillator kernel becomes oscillatory,   
in the same way as without turbulence.
The critical ${\rm Re}_{\rm p}$ where this happens is much smaller
than the particle Reynolds numbers where bistability~\cite{coletti2023} or fluttering~\cite{auguste2013falling,pesavento2004falling} is observed. 
We remark that a preliminary analysis 
suggests that 
our model may also explain the bistable angular dynamics observed in 
Ref~\cite{coletti2023}.
Also, dealing with more complex shapes is an important challenge. Ice crystals, for example, may be hollow or lack fore-aft symmetry~\cite{Heymsfield_1973}. This may give rise to additional torques~\cite{Candelier_Mehlig_2016,Roy_2019}.

In summary, our experiments and model calculations show that particle inertia has a strong effect on the angular dynamics of atmospheric particles, generally enhancing the orientation fluctuations of settling atmospheric particles, not only in still but also in turbulent air.
This causes increased settling velocities and lateral drift, in contrast to the drift-free pattern observed for steadily settling particles in liquids.
Orientation fluctuations also affect the rate at which non-spherical particles collide~\cite{pumir2016collisional} or fragment, a process important for secondary ice particle production~\cite{Korolev_Leisner_2020,Bhowmick_GRL_2020}.
In addition, fluctuations in the orientation of ice crystals affect the radiative properties of ice-laden clouds, for example, by reducing cloud albedo when solar radiation is parallel to gravity.

\emph{Conclusions.}-- We identified the key importance of particle inertia for the motion of non-spherical atmospheric particles.
Our results, made possible by the concurrent development of a unique experimental setup and by a reliable modelling strategy, show that heavy spheroids settling in air at ${\rm Re}_{\rm p} \sim 1-30$ -- values typical for atmospheric particles -- approach their stable orientation through decaying oscillations. We demonstrated that this behavior is a consequence of particle inertia. This physical effect must therefore be accounted for in models of important atmospheric processes, such as
the radiative properties and evolution of ice-laden clouds, and  residence times and dispersion ranges of volcanic ash or microplastics in the atmosphere.

\begin{acknowledgments}
TB was funded by the German Research Foundation (DFG) Walter Benjamin Position (project no.~463393443).
JG was supported by funding from the European Union Horizon 2020 Research and Innovation Programme under the Marie Sklodowska-Curie Actions, Grant Agreement No.675675.
KG was supported by a grant from Vetenskapsr\aa{}det  (no.~2018-03974). BM was supported by Vetenskapsr\aa{}det (grant no.~2021-4452), and acknowledges a  Mary Shepard B. Upson Visiting Professorship with the Sibley School of Mechanical and Aerospace Engineering at Cornell. Statistical-model simulations were performed on resources provided by the Swedish National Infrastructure for Computing (SNIC). This research was also supported in part by the National Science Foundation under Grant No. NSF PHY-1748958. The G\"ottingen turret was manufactured with the  support from the Mechanical Workshop and Electronic Workshop of the Max Planck Institute for Dynamics and Self-Organisation. We thank J.L. Pierson for pointing out Refs.~\cite{Newsom_Bruce_1994,newsom1998orientational} to us.
\end{acknowledgments}

\vfill \eject
\begin{center}
 \textbf{Appendix~A} \textit{Model for Re$_{\rm p}$ up to $\sim 30$}
 \end{center}
 \mbox{}\\[-7mm]
The dynamics of a settling
particle is determined by Newton's equations for translation and rotation:
\begin{subequations}
\label{eq:eqm}
\begin{align}
\label{eq:eqm_translational}
\tfrac{{\rm d}}{{\rm d}t}{\ve x}= \ve v\,, \,\, m_{\rm p}\tfrac{{\rm d}}{{\rm d}t}{\ve v} = \ve F_{\rm h} + m_{\rm p} \ve g\,,\,\,\\
\tfrac{{\rm d}}{{\rm d}t}{{\ve n}}= \ve \omega \wedge {\ve n}\,,\,\,
\tfrac{{\rm d}}{{\rm d} t}
\big[{\ma J_{\rm p}}({\ve n}) \ve \omega\big]  = \ve T_{\rm h}  \,.
\label{eq:eqm_rotational}
\end{align}
\end{subequations}
Here $\ve x$ is the particle position, $\ve v$ its velocity, $\ve n$ is a unit vector parallel to the symmetry-axis of the particle, and $\ve \omega$ is the angular velocity of the particle. Its mass  is $m_{\rm p}$, and  $\ma J_{\rm p}$ is the particle-inertia tensor~\cite{Kim_Karrila_1991}. The gravitational acceleration is denoted by $\ve g$.

The main difficulty is to determine the hydrodynamic force $\ve F_{\rm h}$ and torque $\ve T_{\rm h}$.  For ${\rm Re}_{\rm p}\ll 1$, they can be determined in perturbation theory~\cite{Brenner_1961,Cox_1965,Khayat_Cox_1989,Dabade_2015}. For larger ${\rm Re}_{\rm p}$ -- as in the experiment -- one can parameterise forces and torques on a spheroid in uniform flow using
ab-initio computer simulations~\cite{ouchene2020numerical,frohlich2020correlations}.
The conclusion is generally that force and torque can be parameterised by introducing
empirically determined correction factors in the perturbative equations of motion. Here we use
\begin{align}
\label{eq:para}
 \ve F_{\rm h}& = \ve F_{\rm h}^{(0)} +C_F  \ve F_{\rm h}^{(1)} \,,\quad
  \ve T_{\rm h} = \ve T_{\rm h}^{(0)} +C_T  \ve T_{\rm h}^{(1)}
\end{align}
with correction factors   $C_F({\rm Re}_{\rm p},\asp)$ and  $C_T({\rm Re}_{\rm p},\asp)$.
$ \ve F_{\rm h}^{(0)}=-{\tfrac{m_{\rm p}}{\tau_{\rm p}} } \ma A\ve v$ and  $\ve T_{\rm h}^{(0)} =-{\tfrac{m_{\rm p}}{\tau_{\rm p}}}\ma C \ve \omega$ are Stokes force and torque
in a quiescent fluid \cite{Kim_Karrila_1991,Happel_Brenner_1983},  with  particle response time $\tau_{\rm p} = \tfrac{2\rho_{\rm p}}{9\rho_{\rm f}}\tfrac{a_\parallel a_\perp}{\nu}$
and resistance tensors $\ma A(\ve n,\lambda)$ and $\ma C(\ve n,\lambda)$ (Eqs.~(4,7) in \cite{Gustavsson_2019}).
The ${\rm Re}_{\rm p}$-corrections are
$\ve F_{\rm h}^{(1)} = -\tfrac{m_{\rm p}}{\tau_{\rm p}}\tfrac{3}{16}\tfrac{a_\perp v}{\nu} (3\ma A - \ma I \hat{\ve v}\cdot\ma A\hat{\ve v}){\ma A} {\ve v}$~\cite{Brenner_1961}, where  $\ma I$ is the unit matrix, $\hat{\ve v} = \ve v/v$,
$v= |\ve v|$, and $\ve T_{\rm h}^{(1)}=F(\asp) \tfrac{m_{\rm p}}{6\pi} \tfrac{{a}^3v^2}{a_\perp \nu} (\ve n\cdot \hat {\ve v})(\ve n \wedge\hat {\ve v})/{\tau_{\rm p}}$~\cite{Cox_1965,Khayat_Cox_1989}, with $a = {\rm max}( a_\|, a_\perp )$.
For spheroids, the shape factor $F(\asp)$ is given in Eqs.~(4.1,4.2) of  Ref.~\cite{Dabade_2015}.
For $C_F\!=\!C_T\!=\!1$, Eq.~(\ref{eq:para}) simplify to known expressions for ${\rm Re}_{\rm p}\ll1$~\cite{Gustavsson_2021}.

\begin{figure}[t]
\centering
\includegraphics[width=\columnwidth]{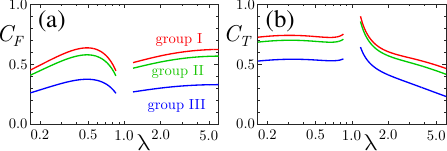}
\caption{\label{fig:CFCT}
Empirical coefficients $C_F$ and $C_T$ in Eq.~(\ref{eq:para}) for the parameters in Fig.~\ref{fig:Fig2}, obtained as described in Appendix~A.
(a)~Force coefficient $C_F$ as a function of aspect ratio $\lambda$. The groups refer to Table~\ref{tab:sd}. (b)~Torque coefficient $C_T$.
}
\end{figure}

Eq.~(\ref{eq:para})
yields good results for $ \ve F_{\rm h}$ up to ${\rm Re}_{\rm p} \sim 100$,
for prolate particles with moderate $\lambda$~\cite{frohlich2020correlations}, and there is
similar qualitative agreement for oblate particles~\cite{ouchene2020numerical}.
 We determined  $C_F({\rm Re}_{\rm p},\asp)$ from interpolations of ab-initio simulation results for fixed ${\rm Re}_{\rm p}$ and $\asp$ for oblate~\cite{ouchene2020numerical}  and prolate~\cite{frohlich2020correlations} spheroids as follows. The full model, Eqs.~(\ref{eq:eqm},\ref{eq:para}) has 11 dimensions. Since $\ve x$ is slaved to the other variables, it is sufficient to
analyse the eight-dimensional system for $\ve v$, $\ve \omega$, and $\ve n$.
In the experiments, the particles settled with speeds close to the steady-state settling speed. Therefore it suffices
to evaluate $C_F$  at the steady state
\begin{equation}
\ve v^\ast =  v_g^\ast\hat{\ve g}\,,\quad \ve \omega^\ast = 0\,,\quad \varphi^\ast = 0,\tfrac{\pi}{2}\,.
\label{eq:FP}
 \end{equation}
Rotational symmetry dictates that the polar angle $\theta^\ast$ can take any value.
Solving Eqs.~(\ref{eq:eqm},\ref{eq:para}) with a given (not yet known) value of $C_F$, we find
\begin{align}
    v_g^\ast=\tfrac{4\nu}{3A^{(g)} a_\perp C_{\rm F}}\Big(\sqrt{1+\tfrac{3}{2}C_{\rm F}\tfrac{a_\perp g \tau_{\rm p}}{\nu}}-1\Big)\,.
\label{eq:vgstar}
\end{align}
Here $A^{(g)}$ is the component of the translational resistance tensor $\ma A$ in the direction of gravity, for a particle falling  with its steady-state orientation $\varphi^\ast$.
We used Eq.~(\ref{eq:vgstar}) to evaluate the Reynolds number, ${\rm Re}_{\rm p}={a}v_g^\ast(C_F)/\nu$, with $a = {\rm max}\{a_\perp,a_\parallel\}$.
Since $C_F$ depends on ${\rm Re}_{\rm p}$, and ${\rm Re}_{\rm p}$ in turn depends on $C_F$ through $v_g^\ast(C_F)$, we solved the resulting implicit equation numerically
  to find the desired value of $C_F$. The results are shown in Fig.~\ref{fig:CFCT}(a).

We determined $C_T({\rm Re}_{\rm p},\asp)$ by interpolating the data in  Refs.~\cite{ouchene2020numerical,frohlich2020correlations}, using ${\rm Re}_{\rm p}$
  determined as described above. The results are shown in Fig.~\ref{fig:CFCT}({b}), they are consistent with the {\em ab-initio} simulations from Ref.~\cite{Jiang_2021}.

It is not guaranteed that the model works outside the tested parameter range, for example for very thin disks, or for very nearly spherical particles.
Therefore we do not report numerical values for $\lambda$ close to unity in Figs.~\ref{fig:Fig2} and~\ref{fig:CFCT}.
As mentioned above, we determined the functions $C_F({\rm Re}_{\rm p},\asp)$ and $C_T({\rm Re}_{\rm p},\asp)$
only near the steady state, for small  $\delta\varphi = \varphi-\varphi^\ast$ (where $\varphi$ is the tilt angle,
and $\varphi^\ast= 0,\tfrac{\pi}{2}$ is its steady-state value).
This is sufficient  as long as the dynamics does not depart too far from the
steady state.  Figs.~\ref{fig:Fig2}(a) and S5 show that this works very well.
The small drift  of the angular dynamics  in Fig.~\ref{fig:Fig2}({a})  may be due to inaccuracies in $C_F$ or $C_T$.
The accuracy of the model could be  improved  by introducing correction matrices $\ma C_F$ and $\ma C_T$ in Eq.~(\ref{eq:para}),
instead of scalars, with  elements that depend on $\varphi$, in addition to Re$_p$ and $\lambda$.

\begin{center}
\textbf{Appendix~B} \textit{Fitting the model to experimental data}
\end{center}
 \mbox{}\\[-6mm]
This appendix contains the details needed to reproduce the theoretical  fits in Fig.~\ref{fig:Fig2}.
First, Eq.~(\ref{eq:eqm}) and (\ref{eq:para})
can be solved numerically for any initial condition $\ve v_0, \ve n_0$  and $\ve\omega_0$ (eight parameters).
To reduce the number of parameters, we fitted only the initial tilt-angle, $\varphi$,
its angular velocity, $\dot\varphi$, the initial settling speed $v_g$, and the velocity component $v_\perp$ perpendicular to gravity.
We assumed steady-state values for the remaining parameters.
As a result, the dynamics resides in a plane determined by gravity and the direction of $v_\perp$.
  The red lines in Fig.~\ref{fig:Fig2}(a) and Fig.~S5 in the SM \cite{sm} were obtained in this way. We see that the approximation
  works very well.

  Second, to determine the shaded
  regions
in Fig.~\ref{fig:Fig2}(b), we perturbed the initial angular velocity and settling speed away from the above initial conditions, using typical experimental values
for the particles from Table~\ref{tab:sd}.

Third, the solid lines in the top panel of Fig.~\ref{fig:Fig2}(b) were obtained
using Eq.~(\ref{eq:vgstar}). The solid lines in the middle and bottom panels of Fig.~\ref{fig:Fig2}(b) were determined from linear stability analysis of the
 eight-dimensional dynamics of $\ve v, \ve \omega$, and $\ve n$.
 Linearising the dynamics around the fixed point~(\ref{eq:FP}), we obtained the eigenvalues describing exponential relaxation to the steady-state. Two eigenvalues form a complex pair. The real part gives the decay rate and the imaginary part gives $2\pi$ times the frequency.

\begin{center}
\textbf{Appendix~C} \textit{Non-dimensional parameters}
\end{center}
 \mbox{}\\[-7mm]
We non-dimensionalise velocities with  $\tilde v\equiv g \tau_{\rm p}/A^{(g)}$, obtained from Eq.~(\ref{eq:vgstar}) in the limit of small settling speed,  time with the angular-velocity relaxation time
${\tau_\omega\equiv }\tau_{\rm p}J_\perp/(m_{\rm p}C_\perp)$, force with  $m_{\rm p} \tilde v/\tau_\omega$, and torque with $J_\perp /\tau_\omega^2$. Here $J_\perp={m_{\rm p}\tfrac{1+\lambda^2}{5} a_\perp^2}$ is the moment of inertia  of a spheroid around an axis perpendicular to its symmetry axis, and  $C_\perp$ is the
rotational resistance coefficient around this axis~\cite{Gustavsson_2019}. In particular, the steady-state settling speed is non-dimensionalised as $V_g^* = v_g^*/\tilde{v}$.
In these non-dimensional units, all terms in Eq.~(\ref{eq:eqm}) are of order unity, except for $F_{{\rm h}}^{(1)}\sim (\rho_{\rm p}/\rho_{\rm f}) g \, V_{\rm p}/\nu^2$ and $T_{\rm h}^{(1)}\sim (\rho_{\rm p}/\rho_{\rm f})^3(gV_{\rm p}/\nu^2)^2$, where $V_{\rm p} = (4 \pi/3 )a_\perp^2 a_\|$ is the volume of the particle.
So there are two non-dimensional parameters in addition to $\asp= a_\parallel/a_\perp$:
the mass-density ratio $\mathscr{R} = \rho_{\rm p}/\rho_{\rm f}$, and the non-dimensional particle volume $\mathscr{V} = gV_{\rm p}/\nu^2$.
 In the experiment, $\mathscr{R}{ = 996}$, and
 $\mathscr{V} \approx 0.06$ (group I in Table~\ref{tab:sd}), $\mathscr{V} \approx 0.1$ (group II) and $\mathscr{V} \approx 1$ (group III).
In the limit of small settling speeds, the parameters $\mathscr{V}$ and $\mathscr{R}$ are connected to ${\rm Re}_{\rm p} $
by ${\rm Re}_{\rm p} \sim \tfrac{1}{6 \pi } \mathscr{R} \mathscr{V}$, up to a $\lambda$-dependent prefactor.

The overdamped limit of the angular dynamics is obtained when
$\tau_\omega \ll \tau_\varphi$, where $\tau_\varphi \sim \nu/[v_g^\ast]^2$ is the relaxation time of the tilt angle in this limit.
So the overdamped limit  corresponds to $\tau_\omega/\tau_\varphi \sim \mathscr{R}^3 \mathscr{V}^2 \ll 1$.

Willmarth {\em et al.} \cite{willmarth1964steady} quantified
particle inertia for settling disks by the phenomenological parameter  $J^\ast = J_\perp^{{(\rm cyl})}/J_{\rm f}$,
where  $J_\perp^{{({\rm cyl})} } =(\pi/2){\rho_{\rm p}}\lambda a_\perp^5$ is the moment of inertia of a short cylinder around its large axis, and
$J_{\rm f} = \rho_{\rm f} (2 a_\perp)^5$ is proportional to the moment of inertia of a fluid sphere with diameter $2a_\perp$. For oblate spheroids, this expression reduces to $J^\ast \propto \asp \mathscr{R}$.

\begin{center}\textbf{Appendix~D} \textit{Harmonic-oscillator approximation}\end{center}
 \mbox{}\\[-7mm]
The planar dynamics described in Appendix~B can be further simplified if $\delta\varphi$ is so small that its feedback upon the settling
speed $v_g$ can be neglected, and if the transversal velocity is much smaller than the settling velocity.
Then the tilt-angle obeys
 a damped-pendulum equation. In the dimensionless units introduced in Appendix~C it reads:
\begin{align}
\label{eq:model2}
\dot\varphi &= \omega\,,\quad
\dot\omega = -\omega -
\tfrac{1}{2}
{({V}_g^\ast)^2} \, C_T \, h(\asp) \mathscr{R}^3 \mathscr{V}^2 \sin(2\varphi) \,,
\end{align}
The first term on the r.h.s. of the equation for $\omega$ is the rotational damping due to particle inertia. The time scale is chosen so that the prefactor is unity.
The dependence of the function $h(\asp)$
on $\asp$  is shown in Fig.~S6.
For thin disks,   $|h(\asp) | \propto \asp$. In this limit,  the  prefactor in the last term on the r.h.s. of Eq.~(\ref{eq:model2})
evaluates to $|h(\asp) | \mathscr{R}^3 \mathscr{V}^2
\propto J^\ast {\rm Re}_{\rm p}^2$.
This rationalises the use of $J^\ast$ to describe
the effect of fluid inertia on settling disks~\cite{willmarth1964steady}.

Although Eq.~\eqref{eq:model2} is approximate, it captures the main features of the experimentally observed dynamics.
The approach to the steady-state is conveniently analysed by linear stability analysis.
Linearisation of Eq.~(\ref{eq:model2})  yields
the harmonic-oscillator approximation
\begin{align}
\label{eq:linear}
0 & = \delta \ddot \varphi + \delta \dot \varphi +  {({V}_g^\ast)^2}\,C_T\,\mid\! h(\asp) \!\mid  \mathscr{R}^3 \mathscr{V}^2    \, \delta \varphi\,.
\end{align}
In the limit of ${\rm Re}_{\rm p}\ll 1$, a corresponding equation was considered earlier, see for example Eq.~(45) in Ref.~\cite{Gustavsson_2019}.
Eq.~(\ref{eq:linear}) implies that the tilt-angle fluctuation $\delta \varphi$ relaxes to zero
exponentially,
${\delta \varphi \sim a_+ \exp(\decay_+ t) + a_{-} \exp(\decay_{-} t)}$,
with eigenvalues
$\decay_{\pm} = -1/2 \pm 1/2 \sqrt{\Delta}$ and
discriminant $\Delta = 1 -  4 {({V}_g^\ast)^2} C_T {|} h(\asp) {|} \mathscr{R}^3 \mathscr{V}^2$.  The square root is real when $4 ({V}_g^\ast)^2 C_T
{|} h(\asp){|}  \mathscr{R}^3 \mathscr{V}^2 \le 1 $, and purely imaginary otherwise. In the former case,  relaxation towards the steady-state is monotonic.
 In the latter case, it involves oscillations.
We remark that ${|} h(\asp){|} $ takes small values [Fig.~S6]. This is compensated by
$\mathscr{R}^3 \mathscr{V}^2 \gtrsim 10^5$
for the particles in Table~\ref{tab:sd}, with $\mathscr{R}\sim 10^3$.
For these particles, $\Delta$ is negative.
For much smaller ratios of particle-to-fluid mass densities, the condition $\Delta<0$ is harder to satisfy.

The results of the harmonic-oscillator analysis are shown in Fig.~\ref{fig:Fig2} as dashed lines. In the top and middle panels they are indistinguishable from the solid lines.
For the exponential decay rate, by contrast, the harmonic-oscillator approximation differs from the solid lines.
The bifurcation predicted by the harmonic-oscillator analysis is shown in Fig.~\ref{fig:Fig3} (dashed lines).

\begin{center}
\textbf{{Appendix~E}} \textit{{Effect of  turbulence}}
\end{center}
 \mbox{}\\[-6mm]
The effect of turbulence can be modeled by adding a stochastic forcing to the model described in Appendix~A, representing
 the turbulent fluid velocity by a Gaussian random
function   $\ve u(\ve x,t)$ (Eq.~(5)~in Ref.~\cite{Bec_2023}).  Since  $C_F$ and $C_T$ in Eq.~(\ref{eq:para}) are approximated assuming small tilt angles (Appendix~A), we must assume that the turbulent dissipation rate is sufficiently small.
The particle equations
of motion (\ref{eq:eqm},\ref{eq:para}) change
in  the presence of the turbulent flow $\ve u(\ve x,t)$. In the expressions for $\ve F_{\rm h}^{(0,1)}$ and $\ve T_{\rm h}^{(1)}$,
the particle velocity
 $\ve v$ is replaced by the slip velocity $\ve v-\ve u(\ve x,t)$ at the particle position $\ve x$. The second change is that there is an additional
 torque due to the gradients
 of the imposed flow \cite{Jeffery_1922}:
 $\ve T_{{\rm J}}^{(0)}=\frac{m_{\rm p}}{\tau_{\rm p}}[\ma C\ve\Omega(\ve x,t)+\ma H:\ma S(\ve x,t)]$.
 Here $\ve\Omega(\ve x,t)=\tfrac{1}{2}[\ve \nabla \times \ve u(\ve x,t)]$ is half the turbulent vorticity,
 and $\ma S(\ve x,t)$ is the strain-rate matrix,
 the symmetric part of the
 matrix of fluid-velocity
 gradients. The tensors $\ma C$ and $\ma H$ depend on particle shape. For spheroids, they
 are given in Eq.~(7) in Ref.~\cite{Gustavsson_2019}. The colon symbol represents a double contraction of indices~\cite{Happel_Brenner_1983}.

\end{document}

\title{Supplemental Material:\\``Inertia induces strong orientation fluctuations of non-spherical atmospheric particles"}

\author{T. Bhowmick}
\thanks{These authors contributed equally to this work}
\affiliation{Max Planck Institute for Dynamics and Self-Organization, G\"ottingen, D-37077 Germany}
\affiliation{Institute for the Dynamics of Complex Systems, University of G\"ottingen, Friedrich-Hund-Platz 1, G\"ottingen, D-37077 Germany}
\author{J. Seesing}
\thanks{These authors contributed equally to this work}
\affiliation{Max Planck Institute for Dynamics and Self-Organization, G\"ottingen, D-37077 Germany}
\author{K. Gustavsson}
\thanks{These authors contributed equally to this work}
\affiliation{Department of Physics, Gothenburg University, Gothenburg, SE-40530 Sweden}
\author{J. Guettler}
\affiliation{Max Planck Institute for Dynamics and Self-Organization, G\"ottingen, D-37077 Germany}
\author{Y. Wang}
\affiliation{Max Planck Institute for Dynamics and Self-Organization, G\"ottingen, D-37077 Germany}
\author{A. Pumir}
\affiliation{Laboratoire de Physique, ENS de Lyon, Universit\'e de Lyon 1 and CNRS, Lyon, F-69007 France}
\affiliation{Max Planck Institute for Dynamics and Self-Organization, G\"ottingen, D-37077 Germany}
\author{B. Mehlig}
\affiliation{Department of Physics, Gothenburg University, Gothenburg, SE-40530 Sweden}
\author{G. Bagheri}
\email[Corresponding author:\\]{gholamhossein.bagheri@ds.mpg.de}
\affiliation{Max Planck Institute for Dynamics and Self-Organization, G\"ottingen, D-37077 Germany}

%\date{\today}

\begin{abstract}
In this document, we summarise details regarding the experimental setup (section {I}), the model described in appendices A to D in the main text, and model calculations accounting for weak turbulence (section {II}). This document contains six supplemental figures and
one supplemental table.
\end{abstract}


\maketitle

\newpage
\setcounter{figure}{0}
\setcounter{table}{0}
\makeatletter
\renewcommand{\thefigure}{S\arabic{figure}}
\renewcommand{\thetable}{S\arabic{table}}
\renewcommand{\theequation}{S\arabic{equation}}

\onecolumngrid

\vspace*{-1.3cm}
\section*{I. Experiments}
The main part of the experimental setup at the Max Planck Institute for Dynamics and Self-Organization (MPI-DS)  is the G\"ottingen turret [Fig.~1(a) in the main text]. It
consists of an air-filled settling chamber (SC), a particle injector (PI), and
four high-speed cameras synchronised with a high-intensity pulsed LED, used to track the dynamics of the settling particles.
The particles are printed with a 3D printer [\textit{Photonic Professional (GT) User Manual}, Nanoscribe GmbH (2015)], by polymerising acrylic resin IP-S (${\rm CH_{1.72}N_{0.086}O_{0.37}}$)~\cite{Liu_2018} with a
feature size of \SI{1}{\micro\meter} for the  optical setting used [\textit{Photonic Professional (GT) User Manual}, Nanoscribe GmbH (2015)].

The setup allows us to track the three-dimensional translational and rotational motion of individual particles, provided that at least one of the particle dimensions is large enough,  \SI{100}{\micro\meter} or larger. This is necessary because
much smaller particles are too difficult to handle manually.
The cameras track a settling
particle for at least \SI{60}{\milli\meter} with a camera magnification of $1:1$. In this configuration the nominal spatial resolution  is \SI{6.75}{\micro\meter\per\px}.
The following subsections describe the
setup in more detail.

\subsection*{I.1. The G\"ottingen turret}

\begin{figure}[b]
    \centering
    \includegraphics[width=0.94\textwidth]{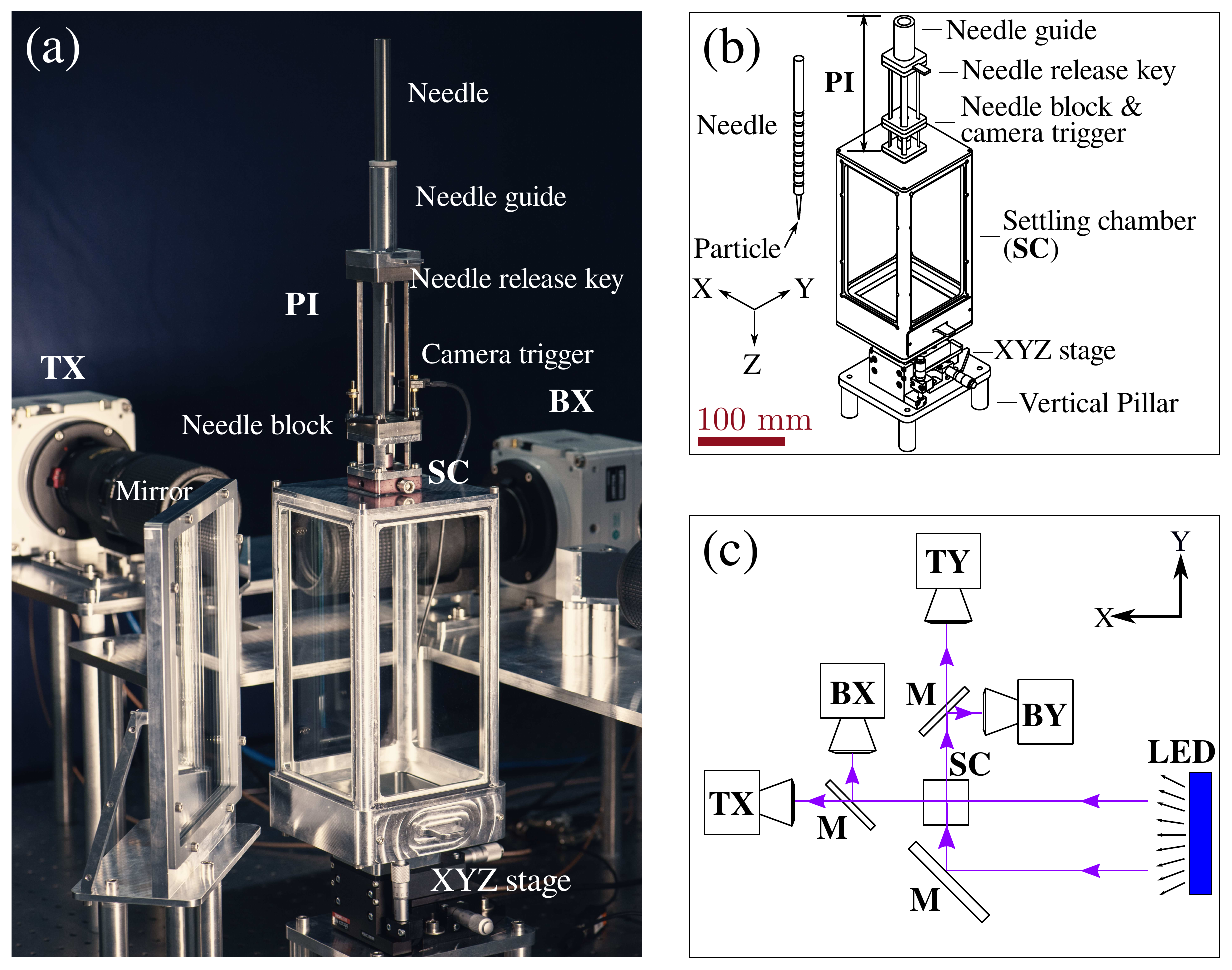}
    \caption{\label{fig:PI}
    (a) Photo of the particle injector (PI) attached to the top cover of the settling chamber (SC). The cameras in the $X$-direction, TX and BX can be seen, as well as the mirror that reflects light through the SC in the $Y$-direction.  (b) CAD drawing of the particle-injector (PI) components, that are installed on the top of the settling chamber (SC). (c) Schematic view of the setup showing the mirror (M) arrangements and illumination/imaging paths.
 }
\end{figure}

Individual particles in the settling chamber are imaged  using four high-speed cameras (Phantom VEO4K 990L, Vision Research),  mounted as shown in Fig.~1(a) in the main text.
The coordinate system in Fig.~1(a) in the main text is defined as follows: $X$ is the direction of the light path from the LED, $Y$ is the horizontal direction orthogonal to $X$, and $Z$ is the direction of gravity.
The two upper cameras (labelled TX and TY in Fig.~1(a) in the main text) cover an observation volume that is \SI{30}{\milli\meter} long in the $Z$ direction at the maximum camera magnification.
The observation volume of the two lower cameras (BX and BY)
also covers \SI{30}{\milli\meter} in the $Z$ direction, and it overlaps slightly with that of the two upper cameras.
The distance between the particle injector and the observation volumes can be adjusted by
an $XYZ$-stage,
and by changing the four aluminium vertical pillars supporting the plate on which the SC is installed (Fig.~\ref{fig:PI}).
Shorter pillars move the observation volume closer to the PI, so that the initial transient dynamics
can be tracked, whereas longer pillars allow imaging the lower part of the SC, to study the steady-state dynamics.

A high-frequency pulsed white LED (LED-Flashlight 300, model number 1103445, LaVision GmbH) was used to illuminate the SC.
Frequency, amplitude, phase, and the duty cycle of the pulsed LED were controlled by a waveform generator (WG).
The exposure time of the cameras and the pulse timing of the LED was synchronised with the strobe signal of one of the cameras.
The pulse duration of the LED determines the effective exposure time of the cameras.
One mirror was used to illuminate the direction perpendicular to the light path of the LED ($Y$ direction), and two additional mirrors were used to reflect the light to the cameras not facing the SC (the cameras BX and BY).

\vspace*{3mm}
\noindent
{\bf I.1.1. Settling chamber (SC)}

\noindent
The air-filled SC is a reinforced steel chamber with a confined volume of $90\times90\times\SI{200}{\milli\meter\cubed}$ in the $X$, $Y$, and $Z$ directions (Fig.~1(a) in the main text and Fig.~\ref{fig:PI}).
Even for the largest particles (group III, $\asp = 0.25$) reported here, the maximum cross-sectional area is only $\sim\SI{0.3}{\square\milli\meter}$, which is less than $0.004\%$ of the SC cross-section.
In other words, the particles were far enough from the walls of the container to neglect additional torques due to wall interactions~\cite{swaminathan2006sedimentation}.
The SC has four glass windows, a detachable top cover with a square opening to  mount the PI, and a detachable bottom drawer to
collect the dropped particles.
With the exception of the opening in the removable top cover, the remaining five edges of the SC are sealed airtight to minimise potential airflow in the SC caused by the motion of the PI,
as well as airflow in the laboratory.
The high-precision $XYZ$-stage is attached beneath the SC, and allows to move the SC with \SI{10}{\micro\meter} spatial resolution in all three directions.
This resolution is a crucial requirement for precise camera calibration.

\vspace*{3mm}
\noindent
{\bf I.1.2. Particle injector (PI)}

\noindent
Since the aim of this study was to investigate the unsteady settling motion of single non-spherical particles, it was necessary  to release
the particles individually, with sufficient control over
their translational and angular velocities.
In particular, one must control the initial direction of motion
so that the particle stays in the focus of all  cameras.
It was also important to ensure that the particles detach from the needle when desired, and that the initial
release speed is neither too small nor too large, so that steady-state motion is reached in the field of view of the cameras. Finally, the flow disturbance in the SC due to particle injection should be minimised.
We designed a particle injection mechanism that meets these requirements, as much as possible: it releases single  particles of different sizes with minimal flow disturbance, and with a high success rate in imaging the particles with all cameras.
Fig.~1(a) in the main text and Fig.~\ref{fig:PI} present the design of the PI.

The experiment was performed by taking a particle from the glass substrate on which it was printed, and placing the particle on the tip of the PI needle (Fig.~\ref{fig:PI}).
The needle is a cylindrical rod with a length of \SI{200}{\milli\meter}, a diameter of \SI{12}{\milli\meter},
and with a conical tip with a end diameter of \SI{1}{\milli\meter} or \SI{1.5}{\milli\meter} for attaching the particles.
The needle was inserted into a hollow needle guide, and locked in place with a release key inserted into one of the needle grooves (Fig.~\ref{fig:PI}).
When the release key was removed, the needle dropped vertically until  stopped by a needle block, where the sudden impulse caused the particle to detach.
Depending on which groove was used, the needle reached a maximal speed of $0.42-\SI{1.5}{\meter\per\second}$, and the particle therefore reached a similar initial speed upon detachment (the locking and eventual release of the needle from higher grooves, at smaller distances between the needle and the needle block, resulted in lower initial particle speeds).
Upon release, the cameras were activated
by an external photoelectric trigger placed
in front of the needle block.

The PI was designed so that  the particles had  negligible lateral velocities, resulting in a  horizontal displacement angle of at most \SI{0.3}{\degree}.
The initial angular velocity
was harder to control, because it is more sensitive to how the particle is attached to the needle tip, as well as the detailed detachment mechanisms.

\vspace*{3mm}
\noindent
{\bf I.1.3.~Cameras}

\noindent Four Phantom VEO4K 990L cameras (Vision Research Inc.) were used to observe the SC. The cameras
were positioned at different heights ($Z$).
The optical axes of the top cameras
 (TX and TY)  were in the
$X$- and $Y$-directions, and the same for the bottom cameras (BX and BY),  see  Fig.~1(a) in the main text.
Only the TX camera received direct light from the LED, the camera BX received light  reflected by a mirror.
The light from the LED was reflected through the SC in the $Y$-direction  using a mirror of $204\times\SI{254}{\milli\meter\squared}$ surface area at a \SI{45}{\degree}-angle facing the LED.
The TY camera received light from this mirror, while the BY camera was illuminated after a \SI{90}{\degree} reflection by another mirror.

All cameras used a Nikon ED AF Micro Nikkor \SI{200}{\milli\meter} $1:4$D lens with a focal length of \SI{200}{\milli\meter} configured at F-stop 22, and a magnification of $1:1$, which corresponds to a image-pixel-size of \SI{6.75}{\micro\meter\per\px}.
The effective exposure time was \SI{2.5}{\micro\second}.
For the range of  measured settling speeds,  $0.2-\SI{1.5}{\meter\per\second}$, the pixel shift
(defined as the exposure time multiplied by the particle velocity, essentially the image streak) was $0.08-\SI{0.56}{\px}$ during the effective exposure time.
The cameras were synchronised as follows.
The camera TX was the leader camera. Its strobe signal was synced to the time codes of the three remaining cameras.
The Phantom camera control (PCC) software was used to configure the role of the cameras and the trigger functions.

With a fixed data throughput of \SI{9}{\giga\px\per\second} [\textit{Phantom VEO-Series Datasheet}, Vision Research, AMETEK (2021)],
increasing the spatial resolution comes at the cost of temporal resolution.
We used the maximum possible resolution (\SI{4096}{\px}) in the $Z$–direction
(the observation volume measured $\SI{27.648}{\milli\meter}$ in $Z$, at $1:1$ magnification, corresponding to images with  $\SI{6.75}{\micro\meter\per\px}$).
The horizontal extent of the observation volume was covered with either \SI{520}{\px} ($\SI{3.51}{\milli\meter}$) at $4000$ frames per second (for spheroids with $\asp = [0.8, 1]$ in group I, and for all  spheroids in groups II and III in Table~1 in the main text, or with \SI{720}{\px} ($\SI{4.86}{\milli\meter}$) at $2932$ frames per second (for spheroids with large aspect ratios $\asp = [0.2, 0.5, 1.25, 2, 5]$ in group I).
Therefore, the horizontal extent of the observation volume was only four times the largest dimension of the $\asp={4}$ particle from Group III, which indicates how small the observation volume had to be
in order to successfully track the  particle.
The lateral displacement of the particle increases when the aspect ratio is much smaller or larger than unity, which affects the probability of observing the particle in focus in all cameras, i.e. the experiment success rate.
The success rate was highest for almost spherical spheroids at around $\SI{34}{\percent}$ and lowest for spheroids with the largest aspect ratios at around $\SI{14}{\percent}$.

\vspace*{3mm}
\noindent
{\bf I.1.4. Illumination}

\noindent
The LED contains 72 white LEDs, summing to an average power consumption of \SI{300}{\watt}. It is operated in the \lq pulse-overdrive\rq\ mode which can emit light at a pulse rate of $0-\SI{20}{\kilo\hertz}$ [\textit{Product Manual, LED-Flashlight 300 white / blue / blue, IP54}, LaVision GmbH (2021)].
The brightness of this LED was larger than \SI{1000000}{\lux} at a distance of \SI{1}{\meter} in the pulse-overdrive mode, and the
duty cycle was between $0-10\%$.

A waveform generator (WG) was used to pulse the LED.
The WG also controlled the pulse rate, amplitude, the offset of the waveform, and the duty cycle of the LED.
The strobe signal from the TX leader camera was
fed into the WG, causing it to generate a synchronisation signal for exposure times of all cameras.
An oscilloscope (OSC) was used to monitor the strobe signals of the cameras, indicating the camera exposure times and phases.

\vspace*{3mm}
\noindent
{\bf I.1.5. Image analysis}

\noindent
The setup was calibrated immediately before or after each day of experiment.
This involved taking images of a transparent and planar calibration target covered with \SI{47250}{} black circular dots simultaneously by all cameras, while the target was placed inside the SC.
The circular dots on the calibration target
had diameter \SI{250}{\micro\meter},  printed with a feature tolerance of less than \SI{0.8}{\micro\meter}, and a dimensional tolerance over the whole calibration target of less than \SI{6}{\micro\meter}.
The calibration target was placed inside the SC at an angle of \SI{45}{\degree} and moved by the $XYZ$-stage in the $XY$-plane (see Fig.~\ref{fig:PI}) at \SI{0.5}{\milli\meter} increments up to \SI{1}{\milli\meter} on either side of the focus plane.

Each camera was first independently calibrated with a camera model \cite{tsai1987versatile}, to minimise the difference between the world coordinates given by the target and their projection recorded during the calibration procedure.
Then, the camera parameters for each camera pair were further optimised by minimising the error between the mean of the three-dimensional world coordinates of the calibration target viewed by both cameras, and the coordinates obtained by stereo triangulation.
In all cases, the root-mean-square error in estimating the $X$, $Y$ and $Z$ world coordinates of dots on the calibration targets was smaller than a pixel, and in most cases below one-third of a pixel.

For each camera, a median image background was calculated using all available frames.
The background was then removed from the images.
A small region around the particle in each frame was then cropped, and the resulting image was independently thresholded.
The centroid position, the longest length, and a fitted ellipse were extracted from the thresholded image using the OpenCV 2 library in Python.
Particle centroids were used to determine the three-dimensional positions of the particles using the stereo-optimised camera model.
The particle velocities were then extracted using a sixth-order finite-difference scheme.
For each spheroid image in a given frame, an ellipse was fitted
(shape and orientation).
For prolate particles, the angle between the major axis of the fitted ellipse
and the gravity defines the tilt-angle.

\vspace*{3mm}
\noindent
{\bf I.1.6. Sensitivity analysis}

\vspace*{2mm}
\noindent
{\em Temperature sensitivity}. To ensure that the air column inside the SC was not affected by the illumination, the change in air temperature in the SC must remain low during operation of the LED.
We monitored the temperature inside the SC with a Humicap\textsuperscript{\textregistered} HMP7
humidity-and-temperature probe from Vaisala.
The temperature sensitivity of the SC was tested in two LED illumination settings: 1) \SI{15}{\second} of illumination following a \SI{105}{\second} of no illumination, and 2) continuous illumination for a total duration of \SI{30}{\minute}, at the highest LED illumination setting of \SI{200000}{\lux} for both the cases.
In the first setting, similar to the way we conducted the experiments, the temperature rise was \SI{0.1}{\celsius} during the illumination time and dropped close to the ambient temperature during the non-illumination time, resulting in a total rise of \SI{0.2}{\celsius} after 30 minutes.
In the second setting, a temperature increase of \SI{1.5}{\celsius} was observed after \SI{30}{\minute}.
The change in air mass density and viscosity due to the \SI{1.5}{\celsius}-temperature rise is negligible, so no buoyancy-induced flow should affect the settling particle.

\vspace*{2mm}
\noindent
{\em Impact of particle injector on the flow inside the SC and setup calibration}.
The movement of the needle before particle injection may result in undesired air currents in the SC.
To quantify this airflow, oil-based smoke was filled into the SC.
We released the PI needle from different locking grooves corresponding to different particle injection speeds, and tracked the smoke movement inside the SC.
We found that the smoke remained stationary when the PI needle was released from most heights, except the highest ones.
When the needle was positioned at the highest available height (second-highest  height), it reached a velocity of \SI{1.5}{\meter\per\second} (\SI{1.32}{\meter\per\second}) before hitting the needle block. In this case,  smoke movement was recorded with a vertical velocity of about \SI{0.07}{\meter\per\second} (\SI{0.01}{\meter\per\second}) near the tip of the needle.

A detailed study of the settling speed of spheres of different volumes $V_p = \SI{1.44e-3}{}$, $\SI{1.70e-3}{}$, $\SI{17.97e-3}{\milli\meter\cubed}$ was carried out to verify the setup.
Table~\ref{tab:Sphere_v_g} compares the observed steady-state settling speed to
the empirical model of Clift \& Gauvin  \cite{Clift_Gauvin_1971}. The relative error
is smaller than  $5\%$.
We note that the stated
relative error of the empirical model
is about $6\%$ at Reynolds numbers below \SI{3e5}{}~\cite{Bagheri_Bonadonna_2016}.

\subsection*{I.2. Printed particles}
The particles were printed using a \lq Photonic Professional GT\rq\ 3D printer from NanoScribe.
This printer polymerises the photoresist liquid with two-photon polymerization (2PP) technique, using \SI{780}{\nano\meter} photons from a 3D Galvo laser scanning system [\textit{Photonic Professional (GT) User Manual}, Nanoscribe GmbH (2015)].
Two printer objectives are available in this setup.
The $63\times$objective was used for high-resolution small-feature printing with an IP-dip 2PP resin on a fused silica substrate with a $200\times\SI{200}{\micro\meter\squared}$ horizontal print field.
The standard hatching distance (horizontal width of each polymerised dot) is \SI{0.2}{\micro\meter}, while the standard slicing distance (vertical thickness of each polymerised dot) is \SI{0.3}{\micro\meter}.
The $25\times$objective was used for medium-resolution printing with an IP-S 2PP resin on an ITO(Indium Tin Oxide)-coated substrate with a $400\times\SI{400}{\micro\meter\squared}$ horizontal print field.
The standard hatching distance for this case was \SI{0.5}{\micro\meter}, while the standard slicing distance was \SI{1}{\micro\meter}.

Once the particles were printed onto the substrate, the un-polymerised portion of the 2PP resin was removed, which is necessary for the particles to be developed.
In this development process, the substrate was immersed together with the particles and the remaining 2PP resin in a SU-8 developer, consisting mainly of 1-methoxy-2-propanol acetate.
This process took about \SI{30}{\minute}, when the un-polymerised resin dissolved in the SU-8.
The SU-8 developer on the substrate and printed particles was then removed by dipping it in isopropanol alcohol for \SI{2}{\minute}.

\begin{figure}[b]
    \centering
    \includegraphics[width=\textwidth]{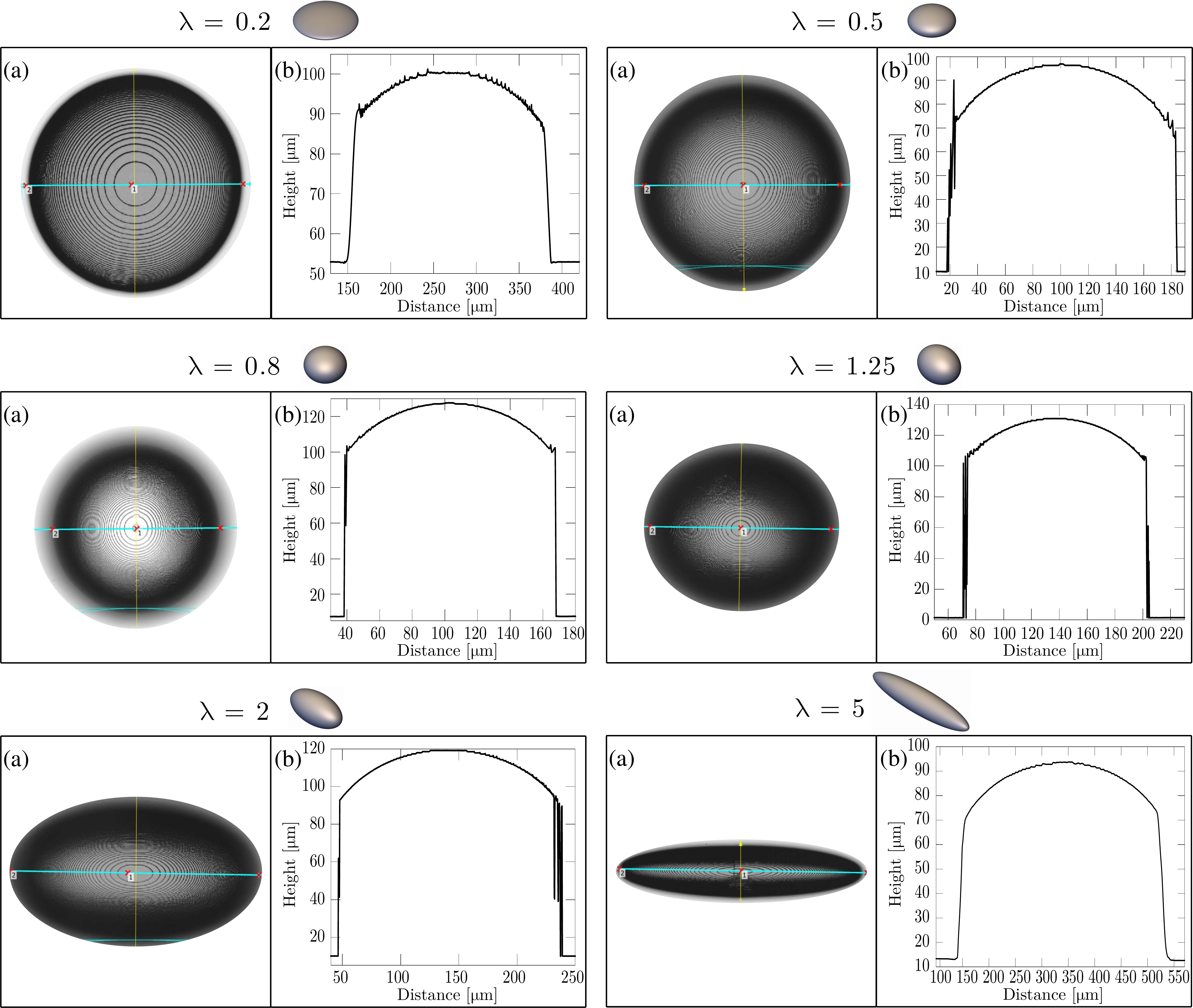}
    \caption{
    Images of 3D scans of the top surface of the printed spheroids with different $\asp$ (Group I in Table 1 in the main text) taken with the Keyence VK-X200K laser microscope at a display resolution of \SI{0.0005}{\micro\meter} (\SI{0.001}{\micro\meter}) for the height (width) measurement [\textit{User’s Manual, 3D Laser Scanning Microscope, VK-X100K/X105/X110, VK-X200K/X210}, Keyence Corporation (2012)]. For each spheroid, the particle surface profile is plotted in (a), and the surface elevation along the horizontal cyan line is shown in panel (b).}
    \label{fig:Spheroid_Dimensions}
\end{figure}

In order to reproduce the increasingly curved outlines
of the slender particles, they were printed using the $25\times$objective in horizontal slices of $0.6$ \SI{}{\micro\meter} thickness.
The printer polymerises the photoresist liquid IP-S to a specific voxel size (of \SI{0.3}{\micro\meter} thickness in our case) using two photons on a conductive ITO-coated substrate.
When the largest axis of the spheroid exceeds \SI{280}{\micro\meter}, the particle was printed in two separate halves, which were then stitched together layer by layer during a later stage of the printing process.
The quality of the resulting particle surface was checked with a Keyence VK-X200K laser microscope, see Fig.~\ref{fig:Spheroid_Dimensions}.
The  resulting unevenness of the particles is in the sub-\SI{}{\micro\meter} range.
We note that it is important to control the particle shape very precisely, because small shape irregularities can result in additional hydrodynamic torques that change the angular dynamics. Weak breaking of fore-aft symmetry, for example, changes the steady alignment angle~\cite{Candelier_Mehlig_2016,Roy_2019}.
How it affects the transient is not known.

In total we printed seven
different spheroids with aspect ratios $\asp = 0.2, 0.5, 0.8, 1, 1.25, 2$ and $5$, keeping
the particle volume unchanged at $V_p = {\SI{1.44e-3}{\milli\meter^3}}$ (group I).
Two additional groups of spheroids (groups II and III) were printed,
with aspect ratios $\asp = 0.25$ ($V_p=$ \SI{2.35e-3}, {\SI{28.28e-3}{\milli\meter^3}}),
and $\asp = 4$ ($V_p=$ \SI{2.08e-3}, {\SI{22.07e-3}{\milli\meter^3}}).
All particle shapes used in this investigation are listed in Table~1 in the main text.
We printed about $60-100$ particles
for each shape.

\subsection*{I.3. Supplemental experimental figures}
\begin{sidewaysfigure}[b]
\centering
\includegraphics[width=\textwidth]{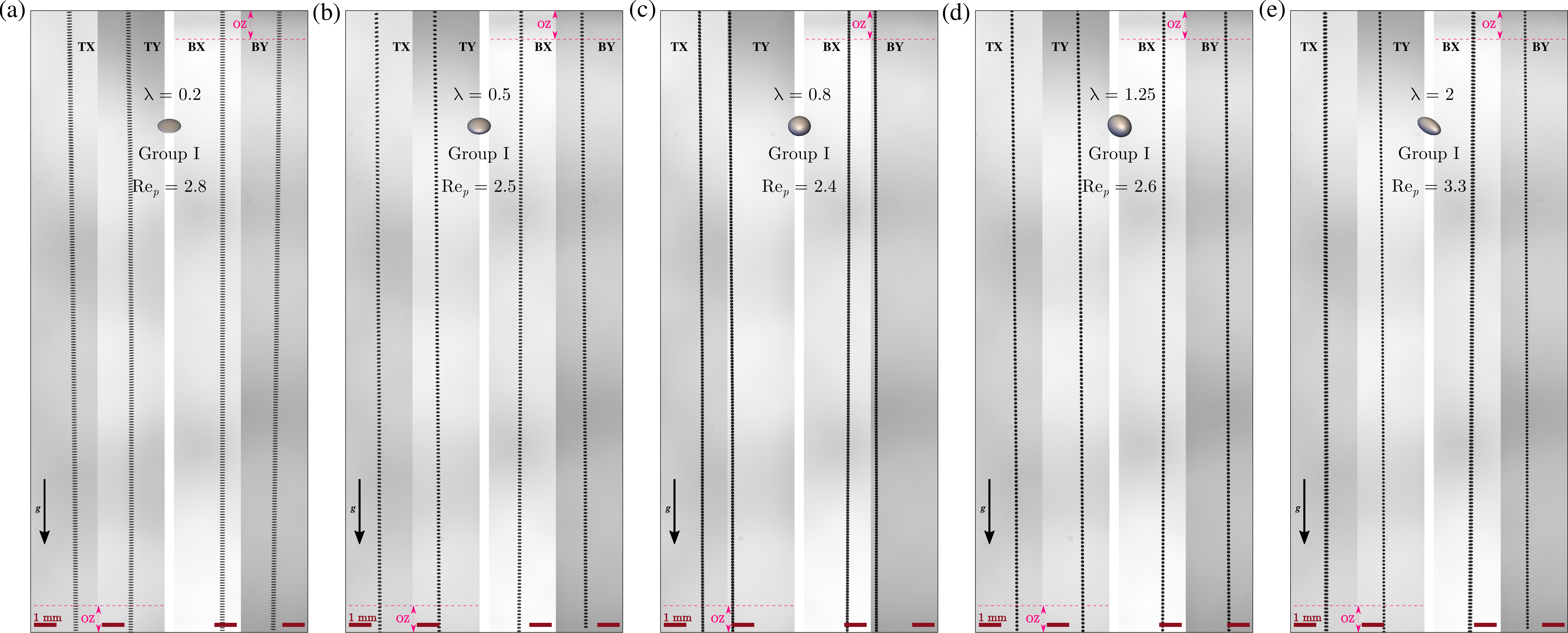}
\caption{
Snapshots of settling spheroids from a single experiment for different particle shapes. (a) $\asp = 0.2$, (b) $\asp = 0.5$, (c) $\asp = 0.8$, (d) $\asp = 1.25$, (e) $\asp = 2$ (all in group I of Table 1 in the main text).
For each shape, data are shown from the top cameras (TX and TY), and bottom cameras (BX and BY) for the
entire observation volume,
i.e, \SI{4096}{\px} or \SI{27.64}{\milli\meter}.
\lq OZ\rq\ refers to the overlapping zone between the
top and bottom
camera pairs, where the particle is observed by all cameras.}
\label{fig:all_overlays_a}
\end{sidewaysfigure}

\begin{sidewaysfigure}[b]
\centering
\includegraphics[width=\textwidth]{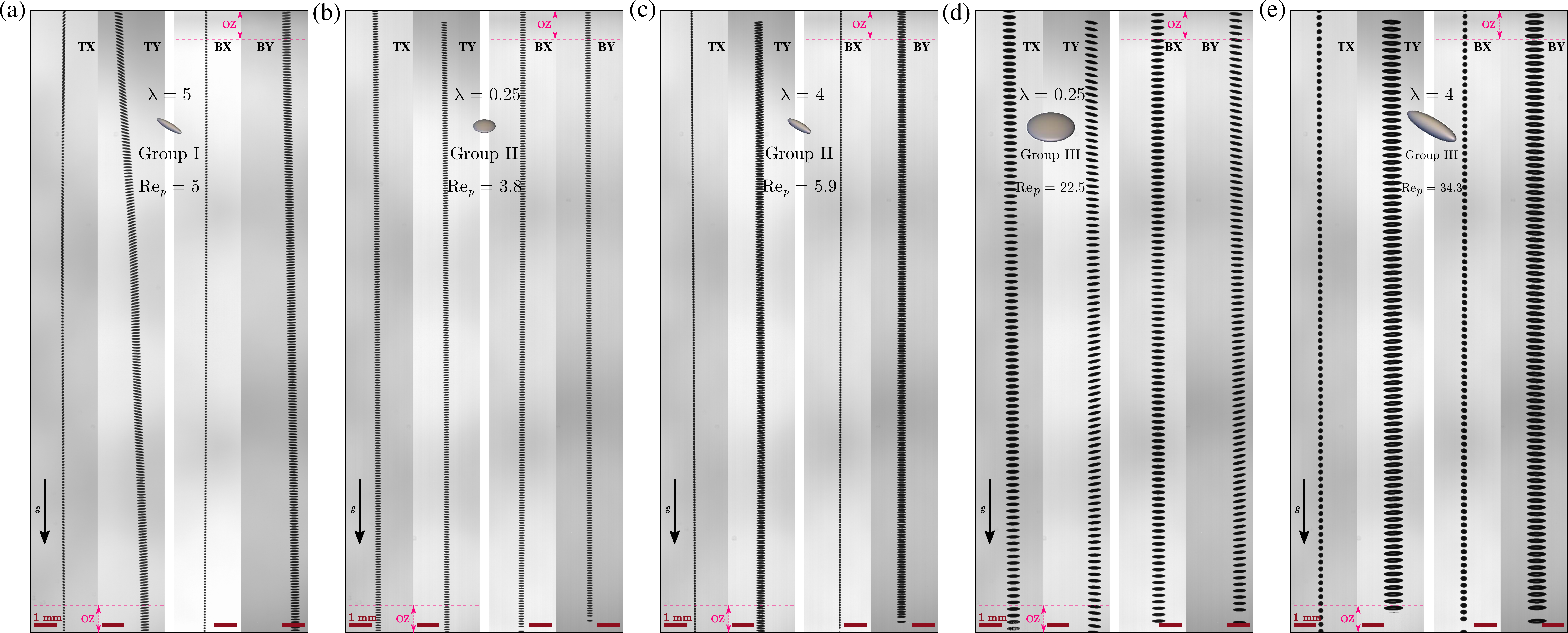}
\caption{
Snapshots of settling spheroids from a single experiment for different shapes (continued from the previous page). (a) $\asp = 5$ spheroid from group I, (b) $\asp = 0.25$, (c) $\asp = 4$ spheroids from group II, (d) $\asp = 0.25$, (e) $\asp = 4$ spheroids from group III of Table 1 in the main text.}
\label{fig:all_overlays_b}
\end{sidewaysfigure}

\begin{sidewaysfigure}[b]
\centering
\includegraphics[width=\textwidth]{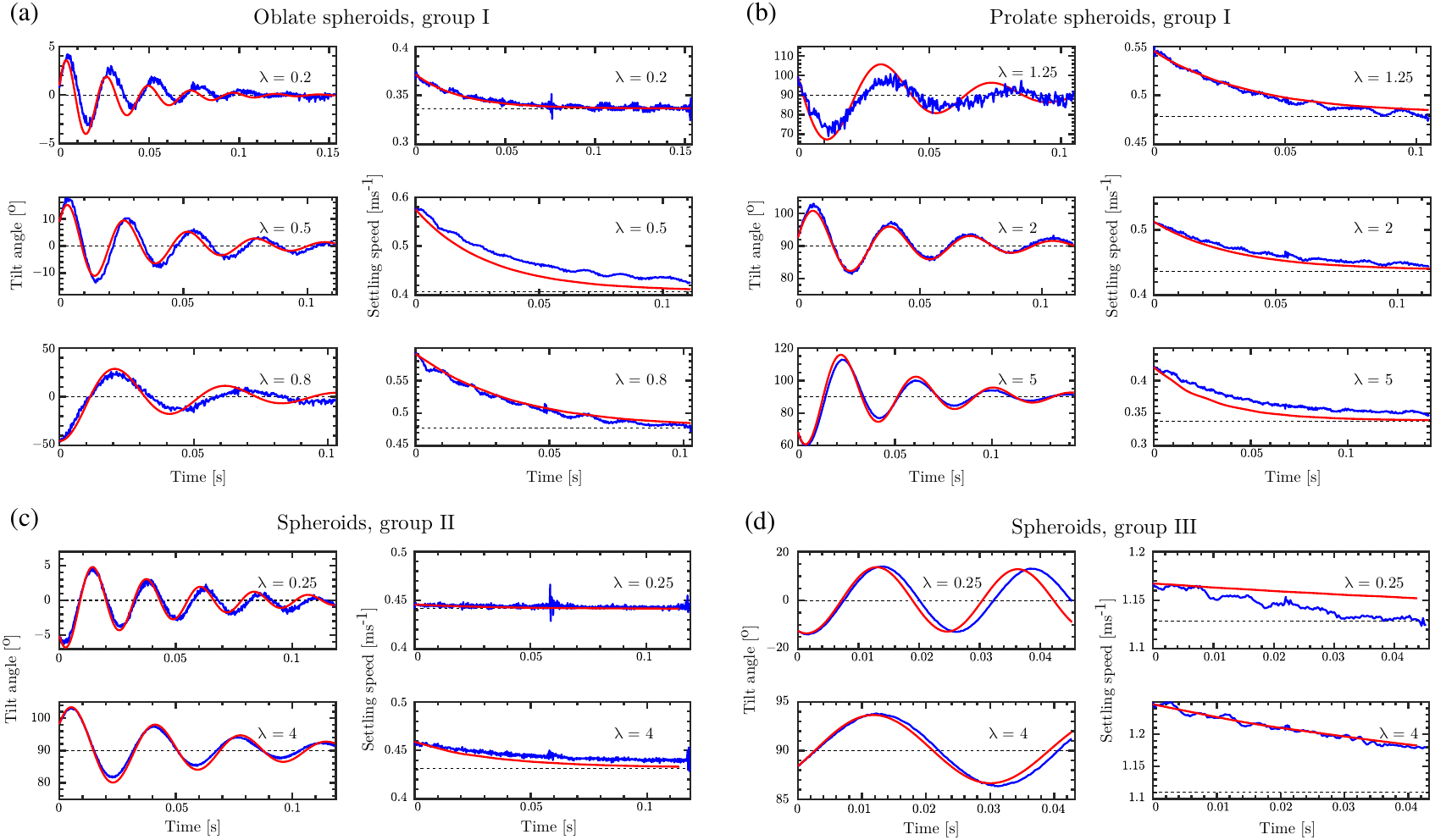}
\caption{
Comparison between experiments and model for the time evolution of tilt-angle and settling speed for the experiments shown in Fig.~\ref{fig:all_overlays_a} and \ref{fig:all_overlays_b}.
(a)~Oblate and (b)~prolate spheroids from group I, (c)~spheroids from group II,
and (d)~spheroids from group III.
Experimental results are shown in blue.
Numerical solutions of the model
are shown in red, matching the  initial conditions to the experiments.
Dashed lines show steady-state tilt-angle and settling speed.
}
\label{fig:exp_vs_theory_all}
\end{sidewaysfigure}

Figs.~\ref{fig:all_overlays_a}-\ref{fig:all_overlays_b} show the recorded images of one experiment for each particle type listed in Table 1 in the main text.
Fig.~\ref{fig:exp_vs_theory_all} shows in blue solid lines the time series of tilt-angle and settling speed post-processed from the experimental data of Figs.~\ref{fig:all_overlays_a} and \ref{fig:all_overlays_b}, together with the model predictions in red solid lines.
The noise seen in the experimental time series for the settling speed
is due to experimental uncertainties of the particle position in the sub-pixel range, caused by mirror distortion, out-of focus imaging, and uneven backlighting between adjacent LED lamps.

\section*{II. Theory}
\subsection*{II.1.~Model for the effect of weak turbulent fluctuations}
In turbulence,
fluid-velocity gradients give rise
to an additional torque, Jeffery's torque \cite{Jeffery_1922}
in the Stokes approximation.
Refs.~\cite{Gustavsson_2019,Gustavsson_2021} used
this model to describe the angular dynamics of small spheroids settling
in a turbulent flow. They accounted for
the inertial torque $\ve T_{\rm h}^{(1)}$ but ignored
$\ve F_{\rm h}^{(1)}$, because its contribution is small in the limit of small ${\rm Re}_{\rm p}$. The results  indicate that particle inertia -- the acceleration terms in Eq.~(1) in Appendix~A -- can have a significant effect on the settling dynamics.  This model  can at best give qualitative insight into the angular
dynamics of atmospheric particles, because it was derived for
${\rm Re}_{\rm p} \ll 1$, while atmospheric particles tend to have
much larger particle Reynolds numbers.

Starting from the improved model described in Appendix~A, the effect of weak turbulence can be modeled  as described in Appendix~E, by adding a stochastic forcing.
 To this end, one represents the turbulent fluid velocity by a Gaussian random
function  $\ve u(\ve x,t)$ with root-mean square velocity $u_{\rm f}$, correlation time $\tau_{\rm f}$, and correlation length $\ell_{\rm f}$~\cite{Bec_2023}.
The particle equations
of motion  -- Eqs. (1) and (2) in Appendix~A -- change
in  the presence of an
imposed external flow $\ve u(\ve x,t)$. In the expressions for $\ve F_{\rm h}^{(0,1)}$ and $\ve T_{\rm h}^{(1)}$,
the particle velocity
 $\ve v$ is replaced by the slip velocity $\ve v-\ve u(\ve x,t)$ at the particle position $\ve x$. The second change is that there is an additional
 torque due to the gradients
 of the imposed flow, computed first
 by \citet{Jeffery_1922}. In dimensional form it reads:
 ${\ve T^{(0)}_{\rm J}}=\frac{m_{\rm p}}{\tau_{\rm p}}[\ma C\ve\Omega(\ve x,t)+\ma H:\ma S(\ve x,t)]$. Here $\ve\Omega(\ve x,t)=\tfrac{1}{2}[\ve \nabla \times \ve u(\ve x,t)]$ is $\tfrac{1}{2}$~times the vorticity of the undisturbed flow,
 and $\ma S(\ve x,t)$ is the strain-rate matrix,
 the symmetric part of the
 matrix of fluid-velocity
 gradients of the undisturbed flow. The tensors $\ma C$ and $\ma H$ depend on particle shape. For spheroids, they
 are given in Eq.~(7) in Ref.~\cite{Gustavsson_2019}. The double contraction represented by the colon symbol is explained in Ref.~\cite{Happel_Brenner_1983}.
Fig.~3(b) in the main text shows results for the r.m.s. of the tilt-angle fluctuations in weak turbulence that were obtained by simulations of the resulting model (symbols).

Also shown in Fig.~3(b)  is the  r.m.s. tilt-angle in the overdamped limit (solid lines).
In this limit, the relaxation time of the angular velocity is much smaller than that of the tilt angle, $\tau_\omega \ll \tau_\varphi$. Here $\tau_\omega=\tau_{\rm p}J_\perp/(m_{\rm p}C_\perp)$ and $\tau_\varphi {\sim} \nu/[v_g^\ast]^2$~\cite{Gustavsson_2019}. In these expressions, $v_g^\ast$ is the steady-state settling velocity, $\nu$ is the kinematic viscosity of air, $m_{\rm p}$ is the mass of the spheroidal particle, $C_\perp$ is the component of $\ma C$ perpendicular to the symmetry axis, and $J_\perp$ is the moment of inertia around an axis perpendicular to the symmetry axis.

The overdamped limit is obtained by neglecting
 translational and angular accelerations. This
corresponds to solving for particle orientation $\ve n$,
 velocity $\ve v$, and angular velocity $\ve \omega$
 by setting force and torque  in Eq.~(1) in the main text to zero.
In a quiescent fluid, the solution is given by the stable fixed point in Eq. (3) of the main text, and the tilt angle $\delta\varphi$ vanishes at this fixed point.

In turbulence, by contrast, turbulent flow-gradients modify
the steady-state orientation, resulting in a small, but non-zero fixed point $\delta\varphi^*$ that depends on the local flow gradients~\cite{Gustavsson_2019,Gustavsson_2021}. As a consequence, components transversal to gravity of the steady-state settling velocity obtain non-zero values. The steady-state angular velocity, however, remains zero~\cite{Gustavsson_2019,Gustavsson_2021}.
An expression for the tilt-angle variance in turbulence in the overdamped
approximation was derived in Ref.~\cite{Gustavsson_2021} (see also references cited there) in the limit of small particle Reynolds number, Re$_p\ll 1$, and disregarding fluid-inertia
correction to translation (the term $\ve F_{\rm h}^{(1)}$ in Eq.~(2) in Appendix~A).
Now we explain how to generalise this calculation to include the fluid-inertia correction to translation, building on the results of Ref.~\cite{Gustavsson_2021}.

When the tilt angle remains small, its angular dynamics  is governed by Eq.~(18) in Ref.~\cite{Gustavsson_2021}.
This dynamics is driven by both the flow gradients and the settling through a matrix $\ma Y$ with components
\begin{align}
Y_{ij}=\frac{1}{\nu}C_T|\mathscr A|A_\parallel A_\perp W_iW_j-O_{ij}-|\Lambda|S_{ij}\,,
\label{eq:Yij}
\end{align}
where $\ve W=\ve v-\ve u$ is the slip velocity, $\ma O$ and $\ma S$ are the antisymmetric and symmetric parts of the flow-gradient matrix, and $\Lambda=(\lambda^2-1)/(\lambda^2+1)$ is a shape factor that depends on the particle aspect ratio $\lambda$.
Moreover, $|\mathscr A|=\frac{5}{6\pi}\frac{{\rm max}(\lambda,1)^3}{\lambda^2+1}|F(\lambda)|J_\perp/(m_{\rm p}A_\parallel A_\perp C_\perp)$, where $A_\parallel$ and $A_\perp$ are the components of $\ma A$ parallel and perpendicular to the symmetry axis of the spheroidal particle. The function $F(\lambda)$ and the coefficients $C_F$ and $C_T$ define the torque model. They are described in Appendix~A in the main text.

Aligning gravity with the $\ve e_1$ direction, the overdamped angular dynamics of a disk-like particle approaches a non-zero stable fixed point (Eq. (S6) in Ref.~\cite{Gustavsson_2021})
\begin{align}
\delta\varphi^*=\pm\frac{\sqrt{Y_{12}^2+Y_{13}^2}}{Y_{11}}\;,\hspace{0.5cm}\theta^*={\rm atan}\left(\frac{Y_{13}}{Y_{12}}\right)\,.
\label{eq:dphiStar}
\end{align}
By solving Eq. (1a) of the main text with zero acceleration, with the orientation given by Eq.~(\ref{eq:dphiStar}), gives the transversal components of the steady-state settling velocity of disk-like particles. In the limit of large enough settling velocity, so that $\delta\varphi$ is small, and the vertical component $W_g$ is much larger than the transversal components as well as the fluid gradients times the Kolmogorov length, we have
\begin{align}
W_g^*&=v_g^*(C_F)\\
W_k^*&=\nu\frac{1+\frac{3}{16}C_F{\rm Re}_{\rm p}(3A_\perp+2A_\parallel)}{1+\frac{3}{8}C_F{\rm Re}_{\rm p}A_\parallel}\frac{A_\parallel-A_\perp}{A_\perp A_\parallel^2C_T|\mathscr A| v_g^*}(O_{1k}-\Lambda S_{1k})
\end{align}
for $k=2,3$,  and where $v_g^*$ is given by Eq.~(4) in the main text.
Using these expressions in Eq.~(\ref{eq:Yij}) allows to evaluate $\delta\varphi^*$ in Eq.~(\ref{eq:dphiStar})
\begin{align}
\delta\varphi^*=\pm\frac{\nu}{C_T|\mathscr A| A_\parallel^2(v_g^*)^2}\frac{1+\frac{3}{16}C_F{\rm Re}_{\rm p}(3A_\perp-A_\parallel)}{1+\frac{3}{8}C_F{\rm Re}_{\rm p}A_\parallel}\sqrt{(O_{12}-\Lambda S_{12})^2+(O_{13}-\Lambda S_{13})^2}\,.
\end{align}
The steady-state tilt-angle for rod-like particles can be derived in a similar manner, following the analysis in Ref.~\cite{Gustavsson_2021}.
Averaging over a Gaussian distribution of fluid gradients gives
\begin{align}
    {\sigma_\varphi^2}=
    \frac{\varepsilon\nu}{({v_g^\ast})^4}\frac{{12}\pi^2}{125}(2+\lambda^2+2\lambda^4)
    \left[
    \frac{{A^{(p)}} C_\perp {m_{\rm p}}}{{A^{(g)}}C_{\rm T}F(\lambda){J_\perp}}
    \frac{1+\tfrac{3}{16}C_{\rm F}{\rm Re}_{\rm p}{(3A^{(p)}-A^{(g)})}}{1+\frac{3}{8}C_{\rm F}{\rm Re}_{\rm p}{A^{(g)}}}
    \right]^2
    {\times\left\{
    \begin{array}{ll}
    2 & \mbox{if }\lambda<1\mbox{ (disk-like)}\cr
    1 & \mbox{if }\lambda>1\mbox{ (rod-like)}
    \end{array}
    \right.
    \,.}
    \label{eq:deltavar}
\end{align}
Here $A^{(g)}=A_\parallel$ and $A^{(p)}=A_\perp$ for disk-like particles ($\lambda<1$) and {\em vice versa} for rod-like particles.
Further, $\varepsilon$ is the turbulent dissipation rate per unit mass.
From Eq.~(\ref{eq:deltavar}), one obtains Eq.~(16) in Ref.~\cite{Gustavsson_2021}
by letting $C_{\rm F}=0$ and $C_{\rm T}=1$. In this limit, the steady-state settling velocity in a quiescent flow, Eq.~(4) in Appendix~A in the main text, simplifies to $v_g^\ast=g\tau_{\rm p}/{A^{(g)}}$.
In Fig.~3(b) in the main text, the results obtained from Eq.~(\ref{eq:deltavar}) are shown as
solid lines.

The tilt-angle variance $\sigma_\varphi^2$ given by  Eq.~(\ref{eq:deltavar}) depends linearly on the energy-dissipation rate
per unit mass $\varepsilon$.
The dependence upon
the  kinematic viscosity $\nu$, by contrast,
is non-linear because
the viscosity is also
contained in the steady-state settling speed $v_g^\ast$,  and in the particle Reynolds number ${\rm Re}_{\rm p}= v_g^\ast a/\nu$.
The parameters $\varepsilon$ and $\nu$ must be mapped
to the statistical-model parameters.
As described in Ref.~\cite{Bec_2023},
we match the Kolmogorov time $\sqrt{\nu/\varepsilon}$ to the time scale of the flow gradients, $(2\langle{\rm tr}(\,\ma S \ma S ^{\sf T})\rangle)^{-1/2}$, and the Kolmogorov length $(\nu^3/\varepsilon)^{1/4}$ to $\sim 0.1$ times the correlation length  $\ell_{\rm f}$ of the random function $\ve u(\ve x,t)$.

Also shown  in Fig.~3(b) is the critical Reynolds number Re$_{\rm c}$ where the discriminant $\Delta$ in the harmonic-oscillator approximation changes sign (see appendix~D in the main text).
From the condition $\Delta=0$, we obtain
\begin{equation}
{\rm Re}_{\rm c} =
{\rm max}(\lambda,1)\big/\big(12\pi A^{(g)}\sqrt{C_T|h(\lambda)|{\mathscr R}}\big)\,.
\label{eq:Rec}
\end{equation}
The function $h(\lambda)$ is shown in Fig.~\ref{fig:h}, and $\mathscr{R}=\rho_{\rm p}/\rho_{\rm f}$ is the mass-density ratio.
Eq.~\ref{eq:Rec} is an implicit condition because $C_T$ depends on ${\rm Re}_{\rm c}$.
This condition was solved to obtain the value in Fig.~3(b), but the scale of ${\rm Re}_{\rm c}$ can be obtained immediately by using $C_T\sim 1$ for the particles considered in this paper.

\subsection*{II.2.~Supplemental theory figure}
Fig.~\ref{fig:h} shows a plot of the function $h(\lambda)$ featuring in the harmonic-oscillator equation, Eq.~(5) in Appendix~D in the main text.

%\bibliography{settling_inertia}
\vfill\eject

\begin{figure}[b]
    \centering
    \begin{tabular}{cc}
        \includegraphics[width=0.65\textwidth]{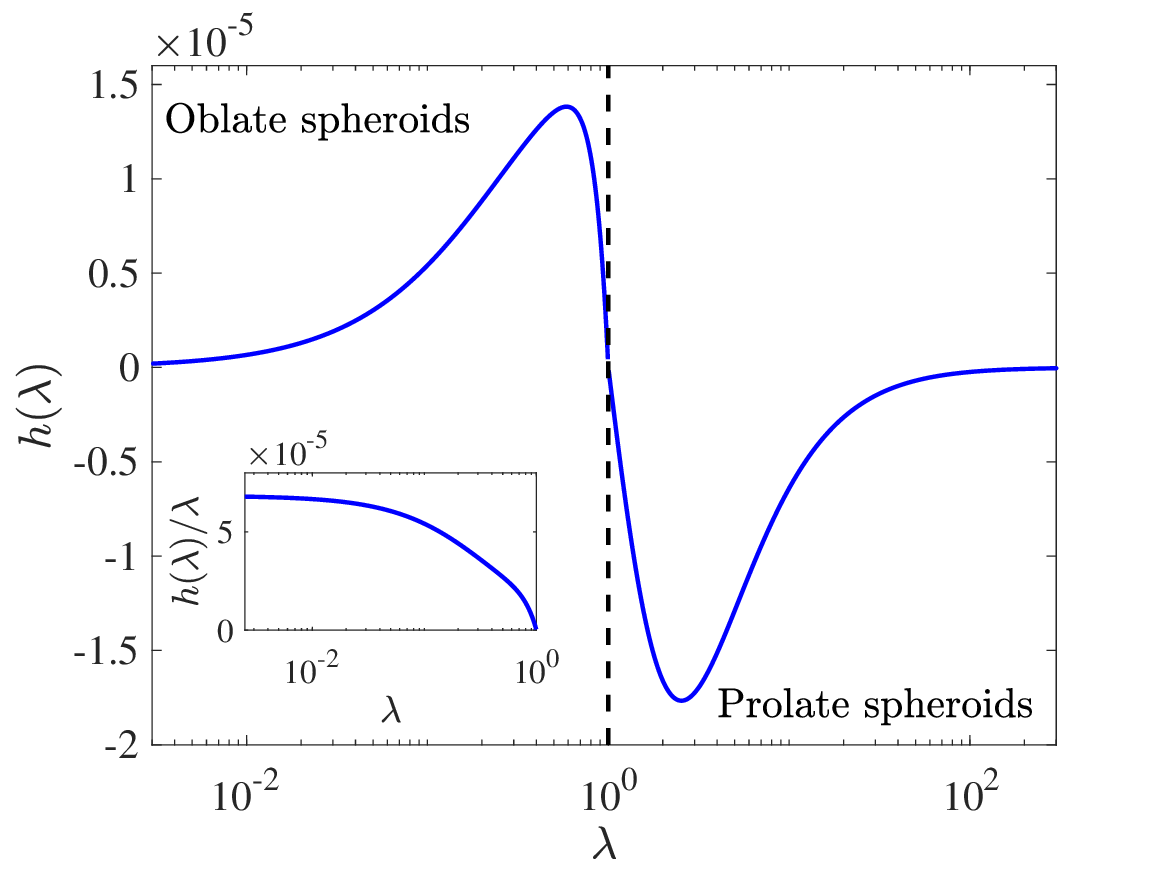}
    \end{tabular}
\caption{
Dependence of the function $h(\asp)$
introduced in connection with Eq.~{(5) in Appendix~D in the main text, as a function of $\lambda$.}
The inset shows that  $h(\asp) /\asp$ approaches a constant $\approx \SI{6.83e-5}{}$ as $\asp \rightarrow 0$.
At large $\asp$, the function $h(\asp)$ behaves as  $ \approx \SI{9.89e-5}{}  {\rm log}(\asp)^2/\asp^2$.}
\label{fig:h}
\end{figure}
\vfill\eject

\begin{table}[p!]
\begin{center}
\begin{minipage}{0.7\textwidth}
\caption{
Settling speed of spheres, based on total number of experiments ${\rm N_{exp}}$ for a sphere with volume $V_p$, diameter $D$, and Reynolds number ${\rm Re}_p$ of the particle.
$v_g({\rm e})$ is the observed settling velocity of the sphere together with the standard error.
$v_g({\rm m})$ is its settling velocity from Clift \& Gauvin (1971) empirical model~\cite{Clift_Gauvin_1971}.
Error correspond to the relative error between $v_g({\rm m})$ and $v_g({\rm e})$.
}
\label{tab:Sphere_v_g}
\centering
\begin{ruledtabular}
\begin{tabular}{ccccccc}
        ${\rm N_{exp}}$ & $V_p$ [\SI{}{{\milli}\meter\cubed}] & $D$ [\SI{}{\micro\meter}] & ${\rm Re}_p$ & $v_g({\rm e})$ [\SI{}{\meter\per\second}] & $v_g({\rm m})$ [\SI{}{\meter\per\second}] & Error [$\%$] \\
    \colrule
        16 & \SI{1.44e-3}{} & \SI{140}{} & \SI{2.24}{} & $\SI{0.4835}{}\pm\SI{0.0001}{}$ & \SI{0.4932}{} & \SI{2.01}{} \\
        9 & \SI{1.70e-3}{} & \SI{148}{} & \SI{2.54}{} & $\SI{0.5191}{}\pm\SI{0.0015}{}$ & \SI{0.5354}{} & \SI{3.15}{} \\
        12 & \SI{17.97e-3}{} & \SI{325}{} & \SI{14.94}{} & $\SI{1.3881}{}\pm\SI{0.0106}{}$ & \SI{1.4578}{} & \SI{5.02}{} \\
\end{tabular}
\end{ruledtabular}
\end{minipage}
\end{center}
\end{table}